\documentclass[a4paper,11pt]{article}
\usepackage{jcappub}
\usepackage{amsmath}
\usepackage{enumerate}
\usepackage{natbib}
\usepackage{url}
\usepackage{xcolor}

\newcommand{\appropto}{\mathrel{\vcenter{
  \offinterlineskip\halign{\hfil$##$\cr
    \propto\cr\noalign{\kern2pt}\sim\cr\noalign{\kern-2pt}}}}}

\title{A model to mini-quench early galaxies by balancing stellar feedback and gas accretion}

\author{Judah Luberto}

\author[]{and Steven R. Furlanetto}

\affiliation{Department of Physics and Astronomy, University of California, \\
475 Portola Plaza, Los Angeles, CA, USA}

\emailAdd{judah@astro.ucla.edu}
\emailAdd{sfurlane@astro.ucla.edu}

\abstract{
The \textit{James Webb Space Telescope} (JWST) has discovered a population of low-mass, early galaxies with very low levels of star formation over tens of Myr. These ``mini-quenched'' galaxies were one of the many surprises from JWST and still require a robust explanation. We seek to explain a few important properties of these galaxies, namely the timescales governing their star formation and quenching. We use a simple model that pairs stellar feedback with the forces binding gas to a $z \sim 6$ galaxy, and with plausible parameter choices, this model produces mini-quenching in galaxies approximately as large as observed mini-quenched galaxies -- up to total stellar masses of $\sim 10^{10} M_{\odot}$. In our model, the ejected interstellar gas forms into a spherical shell that expands outward during quenching. Our novel improvement is allowing this expanding shell to prevent accreting gas from falling in, in contrast to other models whose shells do not interact with accreting gas. Using our model, if a galaxy accretes isotropically, we find that stellar feedback can reproduce observed timescales of mini-quenching with plausible galaxy-formation parameter values. However, at low halo masses, the quenching timescale of our model is too long ($> 100$ Myr). A potential solution for a long quenching time is for the galaxy to accrete via filaments, in which case only a portion of the shell interacts with the accreting gas and the quenching timescale decreases to more accepted values. However, the ram pressure from feedback cannot stop filaments with very small opening angles, as occur in some halo formation models, making mini-quenching impossible. Finally, we use this model to estimate the fraction of galaxies undergoing mini-quenching, finding that it can be large.}


\begin{document}
\maketitle
\flushbottom

\section{Introduction} \label{sec:intro}

Analytic models and simulations predict that when a young galaxy grows during Cosmic Dawn, it will fluctuate between periods of higher and lower rates of star formation \cite{Stinson2007_breathing, Merlin2012, Faucher-Giguere2018, Orr2019}. To gain intuition for why this stochasticity might occur, let us consider a simple argument of timescales: the star formation timescale is tied to the dynamical timescale, which controls the speed at which gas can collapse and form dense clouds. These timescales are $\propto (1 + z)^{-3/2}$, and at early epochs the star-formation timescale becomes less than or comparable to the feedback timescales from those stars (a few tens of Myr for supernova). Short star-formation timescales mean early galaxies could form multiple generations of stars at higher rates before feedback could prevent further star formation, leading to higher net star formation efficiencies. The detailed implications of this trend depend on the details of star formation inside galaxies. With a simple galaxy formation model, \cite{Furlanetto2022} found that stochastic star formation could occur in low mass galaxies ($m_{h} \lesssim 10^{11} M_{\odot}$), while \cite{Dekel2023_FFB} argued that it could occur primarily in massive galaxies (the ``feedback-free'' model). 

While some (simple) models of this stochasticity included phases where the accumulated feedback halted all star formation, these episodes were generally assumed to be oversimplifications. However, recent observations have found a population of early ($z \gtrsim 3$) and low-mass ($m_{*, \rm tot} < 10^{10} M_{\odot}$\footnote{Note that we define observed/total stellar mass as $m_{*, \rm tot}$ instead of the more common $m_{*}$. This is because we later denote $m_{*}$ as the stellar mass produced since the beginning of a single star-formation episode in our model.}) galaxies with zero star formation. The constant number densities of these galaxies across $z \sim 4$--$6$ suggests their star formation is episodic and not permanent \cite{Yang2025, Merlin2025} (leading many to call these galaxies ``mini-quenched''). This mini-quenching occurs over a broad range of galaxy masses. For example, across the varied types of observations (e.g., photometric color cuts \cite{Valentino2023, Alberts2024}, photometric fitting \cite{Merlin2025}, spectroscopic observations \cite{Marchesini2023, Strait2023, Looser2024, Covelo-Paz2026}, and spatially-resolved lensed systems \cite{Hutchinson2025, Messa2025}), many of these galaxies have stellar masses up to $m_{*, \rm tot} \sim 10^{9}$--$10^{10} M_{\odot}$, likely corresponding to relatively large halo masses $m_{h} \sim 10^{11}$--$10^{12} M_{\odot}$. 
Apparently, we should put effort into understanding how this mini-quenching occurs and what requirements it places on models. 

The driving mechanism behind this population of mini-quenched galaxies very likely differs from typical quenched galaxies at lower redshifts. Locally, at around $m_{*, \rm tot} \sim 10^{10} M_{\odot}$, most galaxies transition from the blue type to the red type \cite{Martis2016}, suggesting that late quenching is associated with a mass transition, in contrast to the broad range in which it is seen at high redshifts. Second, at late times quenching persists over very long timescales ($\gtrsim 1$--$2$ Gyr), whereas mini-quenched galaxies do not form stars on the order of a few tens of Myr. Because the latter is much smaller than the age of the Universe even during the Cosmic Dawn, mini-quenching appears to be a \emph{temporary} phenomenon, unlike normal quenching. 

Moreover, the \textit{internal} and \textit{external} quenching mechanisms common to local galaxies are as follows (though our summary is brief and the interested reader should see \cite{DeLucia2025_review} for a review). Internally, high-mass galaxies can host active galactic nuclei (AGN). Accretion onto a central black hole can release enough energy to create massive outflows in high-mass galaxies \cite{Silk&Rees1998, Choi2015} or provide enough heating to prevent star formation in these galaxies \cite{Somerville2008}. Also internally, galaxies could be quenched by their morphological features (e.g., \cite{Martig2009_morphologicalquenching}). High-mass ($m_{*, \rm tot} > 10^{10} M_{\odot}$) \textit{early} galaxies might be quenched by similar processes to local galaxies \cite{Carnall2023_massive, Russell2025, Yang2025}, as large  AGN appear in some early quenched galaxies \cite{Carnall2023_singlegal, Baker2025, Stevenson2025}, and they could even affect star formation in the surrounding environment \cite{Zhu2025}. But signatures of AGN are much harder to observe in low-mass galaxies \cite{Hegde&Wyatt2024, Feltre2016, Gutkin2016, Sanders2023}, even though some low-mass galaxies do appear to host large black holes  (e.g., \cite{Harikane2023_AGN, Juodvbalis2025_AGN}).

Additionally, low-redshift galaxies can be quenched by their external environment. High-mass galaxies tend to live in denser environments, and interactions with the surrounding environment can cause quenching (e.g., \cite{Baldry2004, Balogh2017}). Massive early galaxies are found in a wide range of environments \cite{Binh2025, DeLucia2025, Singh2025}, suggesting that environmental quenching mechanisms could be present in galaxies in the most dense environments. Some lower-mass mini-quenched galaxies have been found in overdense environments as well \cite{Alberts2024, Baker2025_environment, Witten2025}, but there is no compelling evidence for an environmental cause.

Here we focus on a model in which stellar feedback ultimately causes mini-quenching. Previous analytic models so far have treated outflowing gas and inflowing gas as operating independently of each other (e.g., \cite{Dave2012, Dekel2013, Furlanetto2022}). Even in models which track the location of a shell of outflowing gas, the shell is typically assumed not to affect accretion (e.g., \cite{Yamaguchi2023}). However, as long as the outflowing gas is incident on the inflowing gas, it is possible that these gases will counteract each others' motion. Therefore, we create a model in which an outflowing thin shell prevents gas from accreting onto the galaxy, choking off the supply of gas for star formation. This approach should be seen in conjunction with previous attempts to model temporary quenching in early galaxies, either through analytic models (e.g., \cite{Gelli2024, Dekel2023_FFB}) or results from simulations (e.g., \cite{McClymont2025_THESAN, Katz2025, Gelli2020_simulation, Dome2024}).

The purpose of this paper is to present a model which can reproduce important observables of early mini-quenched galaxies using stellar feedback prescriptions. In section \ref{sec:source_model}, we present our galaxy source model. Within this section, we include an implementation of gas accretion by filaments and the motion of the gas leaving the galaxy. In section \ref{sec:approx}, we use this model to calculate an approximate maximum halo mass that may be mini-quenched. In section \ref{sec:quenching_timescale}, we compare our model to a few key observables of mini-quenched galaxies. In section \ref{sec:discussion}, we discuss extensions of our model and implications of mini-quenching in general, and in section \ref{sec:conclusion}, we offer the conclusions of our results.

Unless otherwise specified, throughout this work we use a flat $\Lambda$CDM cosmology with $\Omega_{m} = 0.3111$, $\Omega_{\Lambda} = 0.6889$, $\Omega_{b} = 0.0489$, $h = 0.6766$, consistent with the results of \cite{Planck2020}.

\section{A model of feedback from a source galaxy} \label{sec:source_model}

The purpose of this section is to present our model of mini-quenching in early galaxies, in which stellar feedback removes gas from a galaxy and prevents further gas from accreting onto its inner star-forming region.\footnote{ 
We focus on mechanical effects rather than gas heating (e.g., \cite{Gelli2024}) because of fast cooling timescales at early times (scaling as $\appropto (1+z)^{-3}$).}
The process of quenching a single galaxy in our model is as follows:

\begin{enumerate}
    \item A low-mass, early galaxy has just completed a previous mini-quenched cycle and begins to accrete gas onto its inner region. We assume the gas depleted from the previous mini-quenching period is recycled back onto the galaxy so that the gas mass is proportional to the halo mass: $m_{g} = (\Omega_{b}/\Omega_{m}) m_{h}$.

    \item The galaxy forms stars out of the gas in its reservoir. Given that this galaxy was previously quenched, only the newly-created stars produce feedback.

    \item Meanwhile, the galaxy accretes gas, either isotropically or along filaments. The gas accretion adds to the gas reservoir and may be used to form stars until the galaxy is quenched.

    \item The gas reservoir in the galaxy is pushed outwards by stellar feedback and pulled inwards by the halo's gravitational force and the inward momentum of the accreted gas. We model the gas reservoir as a thin shell at the edge of the star-forming region to test whether feedback can overcome the inward forces and drive the gas outward. 

    \item Once the shell begins to move outward, the galaxy is considered quenched because there is no gas inside the galaxy. We label this the \textit{star-formation} timescale, $t_{\rm SF}$. The galaxy stops forming stars, though feedback continues. The forces on the shell cause the shell to move outwards, which we track using a shell motion equation. 
    
    \item As the shell expands, feedback eventually fades (because star formation has stopped), but the accreting gas continues to provide inward momentum to the shell. Our model assumes that the galaxy remains quenched until the shell falls back onto the galaxy. We label this the \textit{quenching} timescale, $t_{\rm quench}$.
\end{enumerate}
In the following subsections, we describe the components of this picture in detail.

\subsection{The halo potential well}

The first step in creating a galaxy is to consider the halo in which the galaxy lives. In the standard theory of galaxy formation, all galaxies form in dark matter halos. For simplicity, we describe the halos as isothermal spheres whose mass profile is

\begin{equation}
    m_{h}(r) = \frac{2 \sigma^{2} r}{G},
\end{equation}
where $\sigma$ is the velocity dispersion of the halo. The velocity dispersion is the virial/circular velocity at the virial radius, $r_{\rm vir}$:

\begin{equation}
    \sigma^{2} (r_{\rm vir}) \approx \frac{G m_{h}}{r_{\rm vir}}.
\end{equation}
The viral radius is defined as

\begin{equation}
\begin{split}
    r_{\rm vir} (z) & = \left[ \left( \frac{200 G}{\Delta_{h} \Omega_{m}} m_{h} \right)^{1/2} \frac{1}{10 H(z)} \right]^{2/3} \\
    & \sim 3 \mathrm{kpc} \left( \frac{1+z}{7} \right)^{-1} \left( \frac{m_{h}}{10^{8} M_{\odot{}}} \right)^{1/3},\label{eq:rvir}
\end{split}
\end{equation}
where $\Delta_{h} \equiv 200$ is the overdensity of a halo
and $H(z)$ is the Hubble constant across redshift. 

\subsection{A simple galaxy growth model} \label{sec:growth_model}

Our model will trace the interaction of feedback with accreting gas, so we require a prescription for the latter. To begin, we initialize our galaxy to have a gas reservoir with mass $m_{g}= (\Omega_{b}/\Omega_{m}) m_{h}$ available for star formation.
Using the framework and notation from \cite{Furlanetto2017_minimalist}, a dark matter halo which has a mass $m_{h}$ grows approximately at a rate

\begin{equation}
    \dot{m}_{h} = A m_{h} (1 + z)^{5/2}, \label{eq:dmh_dt}
\end{equation}
where $A = 0.03 / \rm Gyr$, which is the average growth rate of halos from \cite{Dekel2013}. The baryons accrete at a rate $(\Omega_{b}/\Omega_{m}) \dot{m}_{h}$, so the gas reservoir grows as
\begin{equation}
    \dot{m}_{g} = \frac{\Omega_{b}}{\Omega_{m}} \dot{m}_{h} - \dot{m}_{*},
\end{equation}
where $\dot{m}_{*}$ is the instantaneous star-formation rate in the galaxy. The gas enters the galaxy with an inflow velocity equal to the virial velocity of the halo, $\sigma$, as supported by simulations \cite{Dekel2009, Goerdt2015}.

We parameterize the star-formation rate by a star-formation efficiency (SFE) per free-fall time, $f_{*}$, such that

\begin{equation}
    \dot{m}_{*} = \frac{f_{*}}{t_{\rm ff}} m_{g}, \label{eq:mstar_dot}
\end{equation}
where $t_{\rm ff}$ is the free-fall time of the gas within the region of star formation. For consistency with previous results (e.g., \cite{Faucher-Giguere2018, Furlanetto2022}), we use the free-fall time of a disk galaxy. We follow the scaling relations from \cite{Faucher-Giguere2018} for the free-fall time, which calculates it at the half-mass radius of a disk inside a NFW halo (e.g., \cite{NFW1996}). The free-fall time is $t_{\rm ff} \approx 0.2 t_{\rm orb} \approx 0.0026 / H(z)$, where $t_{\rm orb}$ is the orbital frequency of the disk. 

As for the SFE, the value of $f_{*}$ for individual galaxies is unknown in (early) galaxy formation, though there is an approximate expected range of $f_{*} \sim [0.001,0.1]$ that is commonly used (e.g., \cite{Furlanetto2022}), with $f_{*} = 0.015$ being the standard local theoretical value \cite{Salim2015, Krumholz2018}. Note that the \textit{instantaneous} SFE is different than the time-averaged SFE, $\tilde{f}_{*} = m_{*, \rm tot} / m_{g}$, which is the ratio of the total mass of stars (including stars formed before the initialization of our model; an observed quantity) and the gas mass.

We also require the dust mass, $m_{\rm D}$, in our model. We assume, again for simplicity, that $D/D_{\odot} = Z/Z_{\odot}$, where $D = m_{\rm D} / m_{g}$, $D_{\odot} = 1/162$, and $Z$ is the metallicity in the galaxy. Our fiducial value for the metallicity is $Z/Z_{\odot} = 10^{-4}$, which is representative of extremely pristine gas (often associated with Population III star formation), though we will find in section \ref{sec:outward_forces} that the forces directly associated with metallicity are subdominant (a metallicity of $Z/Z_{\odot} = 0.1$ changes our results by $\sim 1 \%$).

\subsection{Galaxy radius} \label{sec:radius}

We parameterize the radius of the galaxy as $R_{0} (t) = \kappa r_{\rm vir} (t)$. If a galaxy has a rotationally-supported disk, the edge of the star-forming region would have $\kappa \approx 0.035$, the typical spin parameter of the disk \cite{GalaxyFormation&Evolution2010}. This is the smallest extent our galaxies can have.

Early galaxies can form stars in a much larger region because they may have not formed well-defined disks at this point (e.g., \cite{Danhaive2025}). \cite{Sun2026} created a framework of star formation in a turbulent circumgalactic medium and found that stars can form out to $\kappa \sim 0.4$--$0.5$, though at much smaller rates than will be formed at smaller radii. We approximate the numbers of stars forming at $\kappa > 0.4$ as zero and set a maximum extent of $\kappa = 0.4$.

Throughout the paper, we will use $\kappa = 0.035$ as a fiducial choice. We will find that for isotropic accretion our choice of $\kappa$ does not affect the results, although in section~\ref{sec:mh_max_filaments} we will see that it can affect the results when filamentary accretion is included.

\begin{figure}
    \centering
    \includegraphics[width=0.9\linewidth]{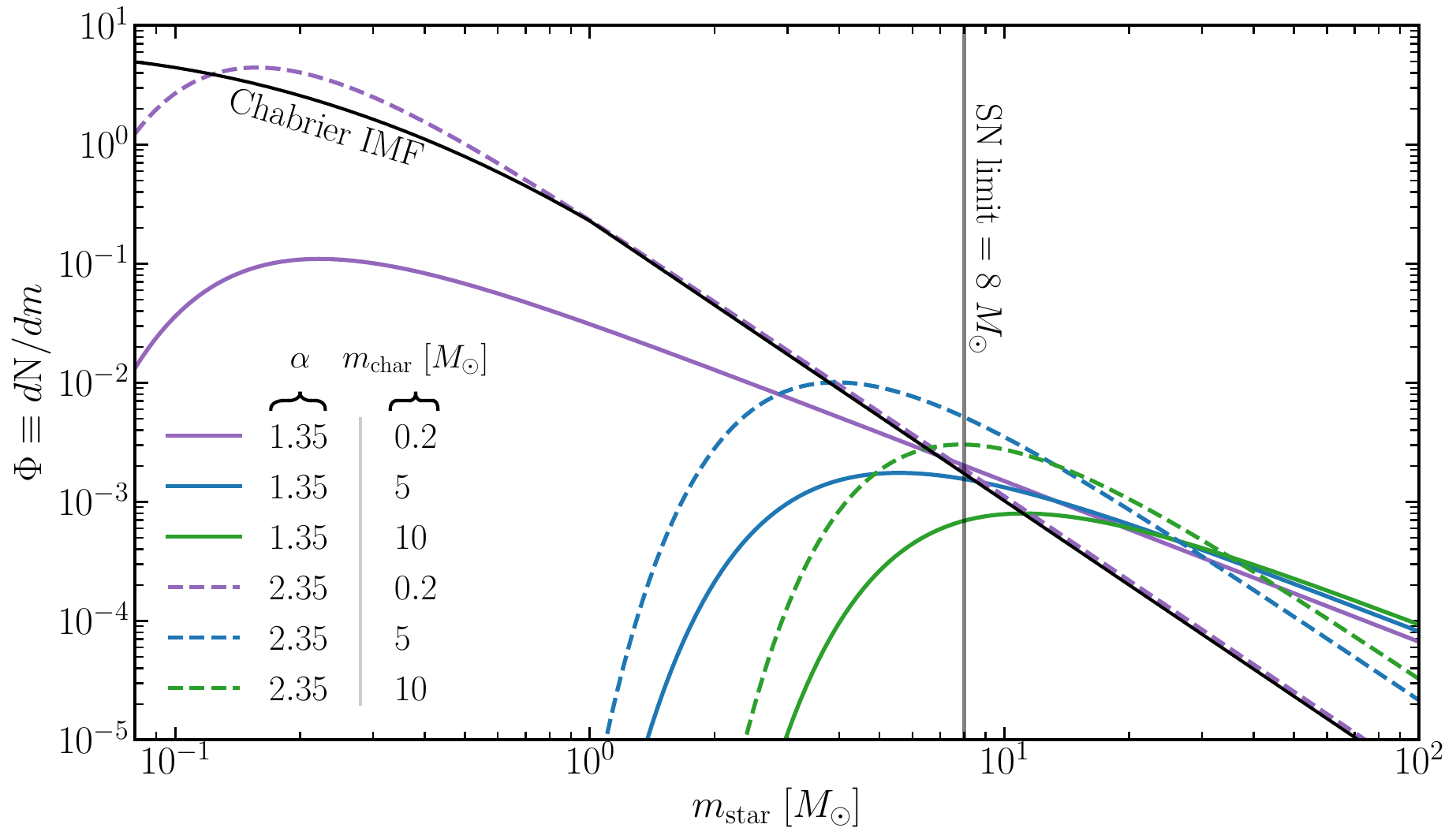}
    \caption{Initial mass functions for a variety of parameter values in equation \ref{eq:IMF}, all with $\beta = 1.6$. We plot the Chabrier IMF as a solid black line and find it is similar to the parameter combination $\alpha = 1.35$, $m_{\rm char} = 0.2 M_{\odot}$, which we define as ``Chabrier-like'' for the remainder of the paper. This IMF is our fiducial choice. However, there exist more top-heavy IMFs with a larger fraction of massive stars -- an important aspect to our model. To gain intuition for the number of supernovae each IMF produces, we plot the minimum mass supernovae can occur at, $m_{\rm star} = 8 M_{\odot}$, as a vertical gray line. We find that the parameter choice $\alpha = 2.35$, $m_{\rm char} = 5 M_{\odot}$ serves to maximize the number of stars in this range (at least for the IMFs plotted).
    }
    \label{fig:IMF}
\end{figure}

\subsection{Initial mass function} \label{sec:IMF}

One of the more important assumptions in our model is the distribution of masses of stars in the galaxy. Because we focus on mini-quenching at early times, we allow our stellar population to follow a more ``top-heavy'' initial mass function (IMF), which is one whose stars exist at higher masses on average, perhaps motivated by its lower metallicity. Given the lack of constraints on the IMF at early times \cite{Glover2026_firststarsreview}, we choose to follow the flexible IMF functional form in \cite{Lazar2022}, which expresses the number of stars per unit stellar mass as 

\begin{equation}
    \Phi (m_{\rm star}) = \frac{dN}{dm_{\rm star}} = \frac{M_{\odot}}{A_{1}} m_{\rm star}^{-\alpha} \exp \left[ - \left( \frac{m_{\rm char}}{m_{\rm star}}  \right)^{\beta} \right], \label{eq:IMF}
\end{equation}
where $A_{1}$ is a normalization constant (defined by $\int m_{\rm star} \Phi dm_{\rm star} = 1 M_{\odot})$. We allow stars to form in a mass range $m_{\rm star} = 0.08$--$100 M_{\odot}$. We choose $\beta = 1.6$ throughout the paper (aligning with, e.g., \cite{Wise2012, Jeon2015, Lazar2022}). The parameter $\beta$ represents the IMF shape at $m_{\rm star} < m_{\rm char}$, which is of lesser importance in our model.

To highlight the flexibility of this functional form, we show in figure \ref{fig:IMF} how it can approximate the well-known Chabrier IMF \cite{Chabrier2003} (solid black line). The dashed purple line uses the functional form of equation~\ref{eq:IMF} with $\alpha =2.35$, $\beta=1.6$, and $m_{\rm char}=0.2 M_{\odot}$. We call this set of parameter values a ``Chabrier-like'' IMF for the rest of the paper. 

The IMF form in equation \ref{eq:IMF} is also flexible enough to create more top-heavy (or bottom-heavy) distributions than the Chabrier IMF. We visualize this effect in figure \ref{fig:IMF} for a number of $\alpha$ and $m_{\rm char}$ parameter combinations. Focusing first on the characteristic mass, we see that increasing the $m_{\rm char}$ value from the Chabrier-like value of $0.2 M_{\odot}$ to $5 M_{\odot}$ (dashed blue line), the IMF distribution shifts quickly toward higher masses.

For our model, the key implication of the IMF is how it affects the amount of stellar feedback, which is driven by massive stars. We might expect a more top-heavy IMF to always produce more feedback. However, more feedback in a stellar population is not a simple mapping to larger $m_{\rm char}$ values: because the average stellar mass increases with $m_{\rm char}$, the number of stars \textit{decreases}. It is possible to increase the average mass \textit{too much}, such that the fraction of stars with masses $> 8 M_{\odot}$ increases, but the total number of supernovae decreases. To understand this complex interplay of parameters, we draw a vertical gray line denoting $8 M_{\odot}$ in figure \ref{fig:IMF} (the minimum stellar mass which can go supernova). Because the IMF drops dramatically at higher star masses, the important quantity to focus on is the value of the IMF at and around $8 M_{\odot}$. Most of the supernovae will occur in this range. For $\alpha = 1.35$, we find that increasing $m_{\rm char}$ from $0.2 M_{\odot}$ to $10 M_{\odot}$ actually decreases the number of stars above $> 8 M_{\odot}$.

Of the IMF parameter sets shown, [$\alpha=2.35$, $m_{\rm char} = 5 M_{\odot}$] has the highest fraction of stars that will go supernova. For the rest of this paper, we will refer to this parameter set as our ``top-heavy'' IMF and use it to show how an IMF more top heavy than the Chabrier IMF will affect the mini-quenching process.

\subsection{Stellar lifetimes}

Because mini-quenching is a time-dependent process, we next need to determine the timescales over which newly-formed stars will drive feedback. We approximate the lifetime of a star as equal to its time on the main sequence. We follow the lead of \cite{Orr2019}, who for simplicity find the stellar lifetimes using a mass-to-light scaling of Population I stars:

\begin{equation}
    m_{\rm star}(t) = \left( \frac{t}{10^{10} \rm yrs} \right)^{-2/5} M_{\odot}, \label{eq:ms_lifetime}
\end{equation}
which also produces produces similar lifetimes as low-metallicity, Population III populations (e.g., \cite{Schaerer2002, Klessen2023}).
Because the smallest star in our model to go supernova is $8 M_{\odot}$, we define the lifetime of that star to be $t_{\rm SN} = 55$ Myr.

\subsection{Radiative feedback from stars}

We next need to determine how feedback from the galaxy's stars drives gas outward (which we model as the force on a thin shell). To begin, we estimate the force produced on a shell from radiation pressure. 
Throughout this section, we refer the interested reader to \cite{Thompson2024_review} for a review.

\subsubsection{UV-driven dusty winds}

One effective source of radiation pressure is the UV emission of massive, hot stars: if the photons are absorbed by dust, and if the dust and the gas are dynamically coupled, the UV photons impart an outward momentum on the gas. We thus begin with the UV luminosity per stellar mass of a population of age $t'$:

\begin{equation}
    l_{\rm UV} (t') = \frac{\int^{x(t')}_{0.08 M_{\odot}} \Phi(m) L_{\rm *, UV} (m) dm}{\int^{100 M_{\odot}}_{0.08 M_{\odot}} m \Phi(m) dm}, \label{eq:luv}
\end{equation}
where $x(t')$ is mass of a star with a lifetime $t'$ (the maximum star mass still alive after time $t'$), $L_{\rm *, UV}(m)  = 4 \pi^{2} R_{*}^{2} (m) B_{\rm UV} (m)$ is the UV luminosity of a single star with mass $m$, $R_{*}$ is the radius of a star of a given mass, and $B_{\rm UV} (m)$ is the Planck function. To convert from stellar mass to temperature, we assume a simple relation of $T_{\rm eff} = 5772\mathrm{K}(m/M_{\odot})^{5/8}$, which matches observations of the solar neighborhood \cite{Eker2018} within $10\%$ for stellar masses $\sim 1$--$20 M_{\odot}$.
The range of relevant UV wavelengths depends on the composition of dust in our galaxies. For example, if early galaxies contain more small dust grains than local galaxies because their dust is created primarily by supernova \cite{Narayanan2025_dustsize}, only the shorter UV wavelengths would be absorbed by dust. But it will turn out that the force from UV-driven winds is sub-dominant even with the full UV wavelength range ($100$--$4000$\r{A}), so we choose to use the entire range for simplicity. With these choices, our model produces about half the number of hydrogen-ionizing photons as a similar prescription with \textsc{starburst99} \cite{Leitherer1999_starburst99}; the main difference is our choice of a Chabrier-like IMF. Our simple top-heavy IMF produces $\sim 20 \%$ more ionizing photons than \textsc{starburst99}. The details of the stellar modeling thus affect our results modestly, with a comparable effect to the IMF choice.

Now, the total luminosity for a galaxy at time $t$ is

\begin{equation}
    L_{\rm UV} (t) = \int^{t}_{0} \dot{m}_{*}(t') l_{\rm UV}(t - t') dt'.
\end{equation}
Finally, assuming that gas and dust are dynamically coupled, the UV-luminosity force on the shell is

\begin{equation}
    F_{\rm UV} = (1 - e^{-\tau_{\rm UV}} + \tau_{\rm IR} ) \frac{L_{\rm UV}}{c},
\end{equation}
where $\tau_{\rm IR, UV} = \kappa_{\rm IR, UV} m_{\rm D} / (4 \pi r^{2})$ is the optical depth of infrared (IR) and UV light, respectively. Here, the IR factor accounts for photons re-emitted by the absorbing dust, which can then scatter again and increase the outward force. For the values of $\kappa_{\rm IR, UV}$, we follow \cite{Thompson2015}: the opacity of dust to UV radiation is $\kappa_{\rm UV} = 10^{3} \rm cm^{2}/g$ and the opacity of dust to IR radiation is 
$\kappa_{\rm IR} = 10^{0.7} \rm cm^{2}/g$, which is
approximately the Rosseland-mean dust opacity over shell temperatures $T \sim 100$--$1000$K. 

\subsubsection{Ly$\alpha$-driven winds}

Early galaxies have less dust and their gas reservoirs are denser, creating conditions where Ly$\alpha$ radiation pressure can be the dominant force from stellar winds. After an ionizing photon is absorbed by the wind, Ly$\alpha$ photons can scatter multiple times within the shell, imparting momentum at each scatter (see \cite{Nebrin2025} for an in-depth calculation of the physics of this process). 

Similar to the results from \cite{Nebrin2025}, the Ly$\alpha$ luminosity per stellar mass formed of age $t'$ is

\begin{equation}
    l_{\rm Ly\alpha} (t') = (1 - f_{\rm esc, LyC}) f_{\rm rec, Ly\alpha} \dot{Q}_{\rm LyC} E_{\rm Ly\alpha},
\end{equation}
where $f_{\rm esc, LyC}$ is the escape fraction of Lyman continuum photons (chosen to be $f_{\rm esc, LyC} = 0.1$), $f_{\rm rec, Ly\alpha}$ is the fraction of recombinations that produce a Ly$\alpha$ photon (chosen to be $f_{\rm rec, Ly\alpha} = 2/3$), $\dot{Q}_{\rm LyC}$ is the rate of production of Lyman continuum photons per solar mass 
(calculated in the same manner as $L_{*, \rm UV}$, but adapted for this quantity), 
and $E_{\rm Ly\alpha}$ is the energy of a Ly$\alpha$ photon. The total Ly$\alpha$ luminosity is

\begin{equation}
    L_{\rm Ly\alpha} (t) = \int^{t}_{0} \dot{m}_{*}(t') l_{\rm Ly\alpha}(t - t') dt'.
\end{equation}

The Ly$\alpha$ force is a function of the Ly$\alpha$ ``force multiplier,'' $\mathrm{M}_{\rm F}$, which reflects the additional force generated by multiple scatterings of each Ly$\alpha$ photon and depends on the hydrogen column density, the dust to gas ratio, and the temperature of the shell (see \cite{Nebrin2025}). 
We approximate the temperature of the shell by assuming half of the dust-driven and Ly$\alpha$ luminosity is deposited as thermal energy and that the gas is cooled by the cooling functions from \cite{Schure2009}, assuming solar abundance, to a minimum temperature of $10^{4} \rm \ K$, below which the cooling times become very long. Our choice of a cooling function derived for higher metallicities predicts faster cooling than the expected lower metallicity population, which results in larger $\rm M_{F}$ values, and therefore more force from Ly$\alpha$ photons. Even with these large $\rm M_{F}$ values, we will find the Ly$\alpha$ forces to be subdominant. With these results in mind, the force from Ly$\alpha$-driven winds is

\begin{equation}
    F_{\rm Ly\alpha} = \mathrm{M}_{\rm F} \frac{L_{\rm Ly\alpha}}{c}.
\end{equation}

\subsection{Feedback from supernovae}

Because supernovae suddenly release an enormous amount of energy into the surrounding medium, we treat their feedback differently than the momentum injection from stellar radiation in our model. 
More specifically, we imagine that supernovae build up a region of hot gas that exerts an outward pressure force against a surrounding (thin) shell \cite{Tegmark1993, Furlanetto2003_winds, Yamaguchi2023}. 

The rate of supernovae at a time $t$ is equal to the rate at which stars above the threshold mass $8 \ M_\odot$ are dying. Given our approximation for the stellar lifetimes in equation~\ref{eq:ms_lifetime}, this can be written as

\begin{equation}
    \dot{N}_{\rm SN}(t) = \int^{t}_{0} {\frac{\dot{m}_\star(t')}{M_\odot}} \left[ \Phi(m_{star}) \frac{d m_{\rm star}}{dt'} \right]_{t-t'} dt', \label{eq:dNsn_dt}
\end{equation}
where the term in square brackets is to be evaluated at the stellar mass whose lifetime is equal to the time interval between the time of a star's formation ($t'$) and the time at which the supernova rate is evaluated ($t$). In appendix \ref{sec:N_analytic}, we analytically calculate $\dot{N}_{\rm SN}$ as a function of $f_{*}$, $m_{h}$, and time of star formation.

Supernovae force the gas shell outwards by supplying energy to the hot gas interior to the shell. Following \cite{Tegmark1993}, we compute the pressure pushing on the shell via

\begin{equation}
    \dot{P} = \frac{L_{\rm p}}{2 \pi r^{3}} - 5 P \frac{v}{r}, \label{eq:dP_dt}
\end{equation}
where $L_{\rm p}$ is the rate of change in energy of the hot gas and $v$ is the velocity of the shell; the last term accounts for the energy loss from $PdV$ work. The supernovae supply energy, but the hot gas also cools (primarily) by Compton cooling (see \cite{Yamaguchi2023} for a discussion on the different forms of cooling):

\begin{equation}
    L_{\rm comp} = - \frac{2 \pi r^{3} P }{t_{\rm comp}}, \label{eq:Lcompton}
\end{equation}
where $t_{\rm comp} = 1.2 \times 10^{8} \ \mathrm{yr} [(1 + z)/10] ^{-4}$ is the Compton cooling time. The total luminosity from the hot gas is then $L_{\rm p} = L_{\rm SN} + L_{\rm comp}$, with $L_{\rm SN} = \epsilon_{k} \dot{N}_{\rm SN} \times 10^{51} \rm erg$. Here, $\epsilon_{k}$ is the fraction of supernova energy released into the wind. We set $\epsilon_{k} = 1$, though we comment on how a change in $\epsilon_{k}$ will change our results in a later section. Finally, the force from supernovae on a shell at radius $r$ is

\begin{equation}
    F_{\rm P} = 4 \pi r^{2} P.
\end{equation}

\subsection{Gas accretion}

The key question for mini-quenching is whether this feedback can halt accretion, choking off the galaxy's gas supply. We now consider how the gas accretes onto the galaxy.

In simple models of early galaxy formation, accretion is often assumed to be isotropic. However, there is support for filamentary accretion at these times: \cite{Mandelker2018} used the C\textsc{olossus} code \cite{Diemer2015_colossus1, Diemer2018_colossus2} to estimate which halos should be fed by filaments as a function of redshift. Using a crude estimate that all halos above a 2$\sigma$ overdensity at any redshift will be fed by filamentary structures, the model estimated that all $z = 6$ halos above a mass $m_{h} \gtrsim 3 \times 10^{9} M_{\odot}$ will be stream fed. According to this argument, most observed early galaxies are fed by filaments of some kind. However, the rapid growth of the nonlinear mass scale makes this expectation highly redshift dependent: at $z \sim 3$, their threshold corresponds to $m_{h} \gtrsim 3 \times 10^{11} M_{\odot}$. While the exact relation between galaxies being fed by streams and their mass/redshift is unknown, it is clear that special attention should be made to the possibility that observed mini-quenched galaxies are supplied cold gas via filaments. We therefore consider both models in which accretion occurs isotropically and those in which accretion occurs along narrow filaments. 

In the latter case, we assume gas filaments subsist down to the star-forming region of the galaxy and are conical, such that the radius of the filament scales linearly with the distance from the center of the galaxy (the same as, e.g., \cite{Mandelker2018}). This assumption is supported by simulations which resolve gas filament accretion in galaxies (e.g., \cite{Dekel2009, vandeVoort2011}). \cite{Ramsoy2021} found that the radius of gas filaments are a few tens of percent of the virial radius (at the virial radius)
for a Milky Way-like progenitor across cosmic time. However, for a broader range of halos 
and filament properties supported by simulations (e.g., \cite{Danovich2012}), 
\cite{Mandelker2018} estimated that at Cosmic Dawn, roughly, the filament radius is $\sim 0.08 r_{\rm vir}$ when evaluated at the virial radius (corresponding to a covering fraction of only $\sim 0.005$). They performed this calculation using an analytic model which assumed cold gas streams are supported by rotation (as they found cold streams supported by zero rotation are inconsistent with cosmological simulations and observations). 

As long as gas filaments are conical, the solid angle of a gas stream as observed from the central galaxy is constant with radius:

\begin{equation}
    \theta_{\rm fil} = \pi \arctan^{2} \left( \frac{r_{\rm fil}}{r_{\rm vir}} \right).
\end{equation}
With this value in mind, the total covering fraction is

\begin{equation}
    f_{\rm cover} = \frac{N_{\rm fil} \theta_{\rm fil}}{4 \pi},
\end{equation}
where $N_{\rm fil}$ is the number of filaments and $\theta_{\rm fil}$ is the solid angle covered by each filament, assuming each filament is the same size. Many cosmological simulations find that galaxies which accrete by filaments do so with one large, primary stream that makes up $\sim 50 \%$ of the accretion rate, and two smaller streams which make up the rest of the accretion rate (e.g., \cite{Danovich2012}), but our model ignores these differences. Therefore, we assume $N_{\rm fil} = 3$ identical streams for our filamentary model (similar to \cite{Mandelker2018}). 

\begin{figure}
    \centering
    \includegraphics[width=0.7\linewidth]{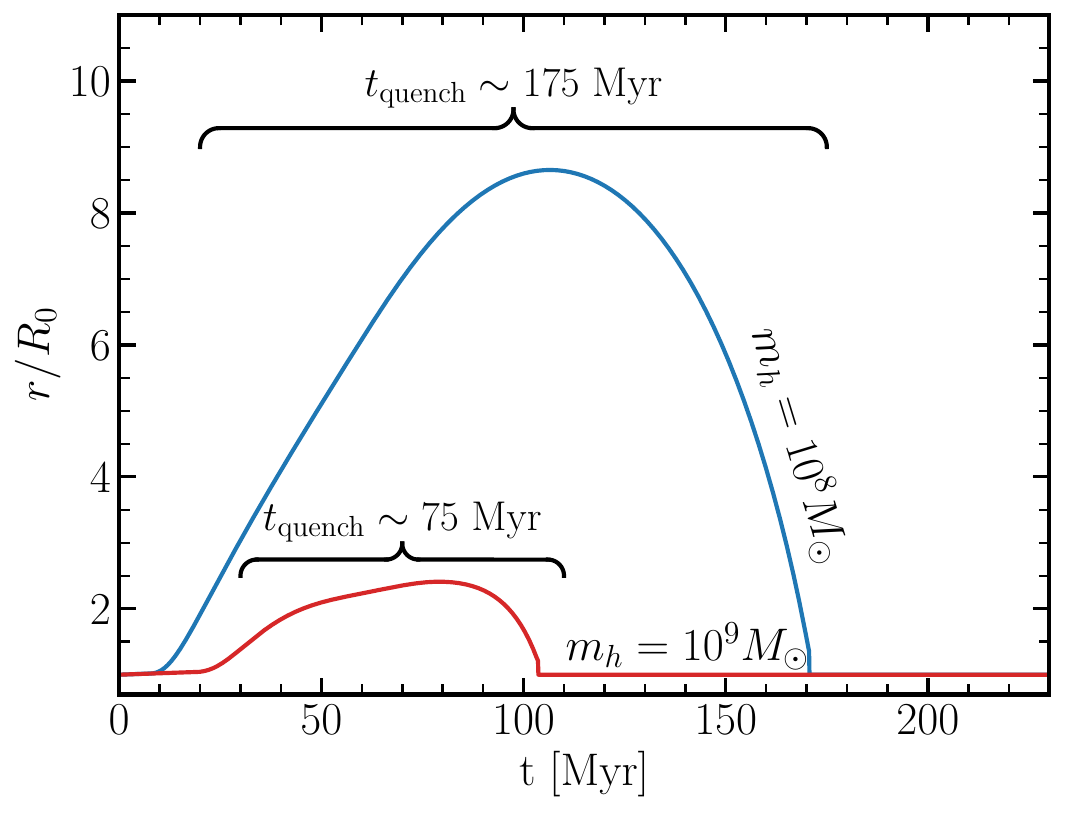}
    \caption{Shell paths of two example galaxies as a function of time, $t$. Each are run with $f_{*} = 0.001$, a Chabrier-like IMF, and $f_{\rm cover} = 1$, and both have star-formation times a few tens of Myr. To display the distance reached by each example, we plot the shell radius, $r(t)$, scaled by the minimum radius, $R_{0}(t)$, on the $y$-axis. The $m_{h} = 10^{8} M_{\odot}$ halo (blue line) reaches a much further scaled radius than the $m_{h} = 10^{9} M_{\odot}$ halo (red line). As a consequence, the lower-mass halo is mini-quenched for a longer period than the higher-mass halo. We illustrate this longer quenching timescale with the brackets above each shell path: the galaxy begins quenching when $r/R_{0} > 1$ and concludes its quenching period when the shell path returns to $r/R_{0} = 1$. 
    Here, the lower-mass halo is quenched for $\sim 2.5 \times$ longer than the higher-mass example.}
    \label{fig:shell_ex}
\end{figure}

\subsection{Motion of the outflowing shell} \label{sec:shell}

To initialize the system, we assume that the galaxy's gas reservoir is in a thin shell at $R_0$, the effective radius of the star-forming region (see section~\ref{sec:radius}). 
However, for our purposes we only need follow the portion of the shell that is incident on the accreting gas, because it determines what actually prevents the gas from accreting. This has an initial mass $m_{\rm sh, f} = f_{\rm cover} m_{\rm g}$. If the shell is outflowing, it gains mass as $\dot{m}_{\rm sh, f} = (\Omega_{b} / \Omega_{m}) \dot{m}_{h}$. Otherwise, the accreting gas gathers into the gas reservoir of the galaxy.

The motion of the outflowing gas shell (incident on the accreting gas) is described via a force equation. Similar equations have been discussed before, both for radiation pressure on dusty shells \cite{Thompson2015, Ferrara2025_feedbackfreepossible, Manzoni2025, Nakazato2025} and supernova-induced pressure \cite{Tegmark1993, Furlanetto2003_winds, Yamaguchi2023}. However, our equation below includes the impact of filaments with the covering fraction of accreting gas:

\begin{equation}
        m_{\rm sh, f} \frac{dv}{dt}  = -\frac{G m_{h} m_{\rm sh, f}}{r^{2}} - \dot{m}_{\rm sh, f} (v_{\rm fil} + v) + (F_{\rm UV} + F_{\rm Ly\alpha} + F_{\rm p}) f_{\rm cover}. \label{eq:shell}
\end{equation}
where $v_{\rm fil} = \sigma$ is the inward velocity of the accreting gas, which is set to be always positive. The behavior of the shell is that when $v$ is positive, the shell moves outwards and the galaxy is considered quenched. In the equation above, the forces $F_{\rm UV}$, $F_{\rm Ly\alpha}$, and $F_{\rm p}$ all scale with $f_{\rm cover}$ because it is the fraction of the total force from the galaxy that is incident on the portion of the shell pushing against accretion. 

We have assumed the gas reservoir is a thin shell at $R_{0}$. Realistically, the gas reservoir will take time to evolve from its diffuse state in a galaxy into a thin shell outside the galaxy's star forming region, but this provides a minimal criterion for an outflow and is reasonable given that low-mass galaxies appear to quench in a few Myr \cite{Gelli2023}. Our parameter $\epsilon_k$ can account for the supernova energy lost in driving the gas from its initial configuration into a shell. We also note that we do not allow the shell to sink to smaller radii to prevent the unphysical result that $r \rightarrow 0$ when $v < 0$. 

Figure~\ref{fig:shell_ex} demonstrates the paths of shells for two example galaxies, both of which form stars with $f_{*} = 0.001$ (and $f_{\rm cover} = 1$). The span between the beginning of shell motion and the end is the quenching timescale, $t_{\rm quench}$. We write above curly brackets the approximate quenching timescales of each model halo as a visualization of the process.

\section{An estimate for the largest mini-quenched galaxy} \label{sec:approx}

In this section, we derive a simple criterion for the limiting halo mass that can undergo mini-quenching by determining the conditions under which outward feedback forces overcome the binding forces of gravity and gas accretion. We explicitly perform this calculation for galaxies at $z \sim 6$, as some of the relevant processes have complex redshift dependence. We show results for other redshifts as well. In later sections, we will see that this simplified estimate of the maximum halo mass matches the results of our full model rather well.

\subsection{Which inward forces dominate?}

To begin, we consider the two inward forces (gravity and accretion) and determine which is more significant. The force of gravity from an isothermal halo on a shell is

\begin{equation}
\begin{split}
    F_{g, \rm max} (R_{0}) & = \frac{2 G m_{h} m_{\rm sh} f_{\rm cover}}{R_{0} r_{\rm vir}} \\ & = 4.2\times10^{32} \mathrm{erg/cm} \left( \frac{m_{h}}{10^{8} M_{\odot}} \right)^{4/3} \left( \frac{\kappa}{0.035} \right)^{-1} \left( \frac{f_{\rm cover}}{1} \right),
\end{split}
\end{equation}
where we have assumed the mass of the shell is equal to the baryonic mass in an average galaxy, $m_{\rm sh} \sim (\Omega_{b}/\Omega_{m}) m_{h}$. The gravitational force depends on the halo mass, which increases throughout the period of star formation;  according to equation \ref{eq:dmh_dt}, it will grow by a factor $\xi = \exp[A(1+z)^{5/2}t_{\rm SF}]$ . At $z = 6$, this factor can reach $\approx 1.5$ for $t_{\rm SF} = 100$~Myr, so we take $\xi = 1.25$ as a representative value and increase the effective gravitational force by a factor $1.25^{4/3}$ following the above mass scaling.

The second inward force is that from gas accreting onto the galaxy, for which

\begin{equation}
\begin{split}
        F_{\rm acc} & = v_{\rm acc} \dot{m}_{g} \\ & \sim v_{\rm vir} \dot{m}_{h} (\Omega_{b} / \Omega_{m}) \\& = 2.9\times10^{31} \mathrm{erg/cm} \left( \frac{m_{h}}{10^{8} M_{\odot}} \right)^{4/3},
\end{split}
\end{equation}
where we have also multiplied the prefactor by $\xi^{4/3}$. The ratio of these two forces is

\begin{equation}
    f_{1} = \frac{F_{g, \rm max}}{F_{\rm acc}} = 9.3 \left( \frac{\kappa}{0.035} \right)^{-1} \left( \frac{f_{\rm cover}}{1} \right),
\end{equation}
which means that the force of gravity is dominant if $\kappa < 0.32$ (and $f_{\rm cover} = 1$). Because the maximum extent of our disk is $\kappa \lesssim 0.4$, we assume below that the force of gravity exceeds the accretion force (for isotropic systems, although section \ref{sec:mh_max_filaments} shows that this may not be true for filamentary accretion).

\begin{figure}
    \centering
    \includegraphics[width=\linewidth]{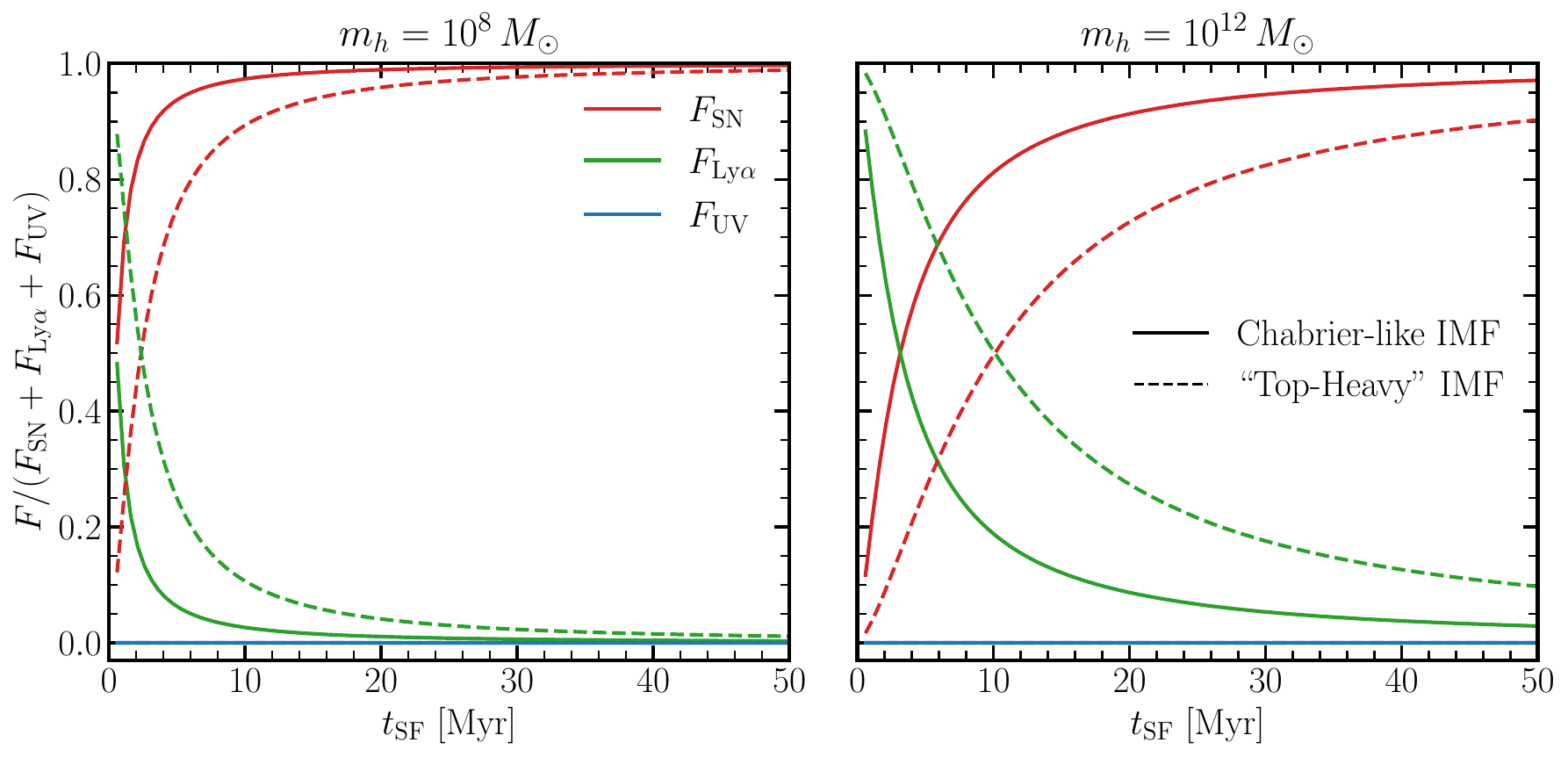}
    \caption{
    Fraction of the outward feedback force provided by each feedback source in our model at $z = 6$. We plot this value for two halo masses as a function of the length of time of star formation, $t_{\rm SF}$, and for our two representative IMFs: a Chabrier-like IMF (solid lines) and a ``top-heavy'' IMF which maximizes the number supernova occurring per unit star formation. In each IMF case, the number of supernovae increases with time (after short timescales, only the rare, extremely-massive stars have gone supernova, while at later times, the more numerous ``lower'' mass stars die). The top-heavy IMF decreases the fraction of force that supernova produce because radiation pressure increases more dramatically than the force from supernova across these IMF ranges.}
    \label{fig:force_comparison}
\end{figure}

\subsection{Which outward forces dominate?} \label{sec:outward_forces}

While it has been found that the Ly$\alpha$ force is the dominant feedback force in the context of radiation pressure on low-metallicity gas \cite{Thompson2024_review, Nebrin2025}, in our model, supernova pressure is dominant on the scales of the galaxy star-forming region. 

In figure \ref{fig:force_comparison}, we plot the fraction that each feedback mechanism contributes to the total outward force of two example galaxies. The ratio of each feedback force to the total force does \textit{not} depend on $f_{*}$ because each force scales linearly with the SFE. However, the ratio does depend on how long each of these galaxies form stars, as shown on the $x$-axis. For a very low-mass galaxy (left panel), the galaxy is dominated by supernova pressure after a few Myr of star formation. However, for a larger halo (right panel), the Ly$\alpha$ force can be strong up to $\sim 10$ Myr. This result suggests that stellar winds might be useful in creating the shell and \textit{causing} mini-quenching on timescales of a few Myr (similar to the results of \cite{Gelli2024}), but supernovae are required to maintain quenching for longer timescales.

To find an analytic expression for the force from supernovae, we assume Compton cooling is subdominant, which (if $f_{*} \sim 0.01)$ agrees with our full equations within $\sim 10 \%$ (see appendix \ref{sec:compton_cooling_approx}). Then, equation \ref{eq:dP_dt} becomes $P_{\rm max} = N_{\rm SN} (t_{\rm SF}) E_{\rm SN} / 2 \pi R_{0}^{3}$, where $N_{\rm SN}(t_{\rm SF})$ is the number of supernovae after a period of star formation. With an expression for $N_{\rm SN}(t_{\rm SF})$ from appendix \ref{sec:N_analytic}, the maximum pressure force from supernova is

\begin{align}
P_{\rm max} & = \ 7.7 \times 10^{-8} \mathrm{erg/cm^{3}} \left( \frac{f_{*}}{0.01} \right) \left( \frac{\kappa}{0.035} \right)^{-3} \\
    & \times 
    \begin{cases}
        \left( t_{\rm SF} / t_{\rm SN} \right)^{1.54} & \text{if } t_{\rm SF} \leq t_{\rm SN}\\
        (t_{\rm SF} - t_{\rm SN})/t_{\rm char} + 1 & \text{if } t_{\rm SF} > t_{\rm SN}.
    \end{cases}
\end{align}
where $t_{\rm char} = 36$ Myr and $t_{\rm SN} = 55$ Myr. As with the inward forces, this expression depends on the halo mass, which will evolve through the star formation episode. The pressure force scales as $\propto N_{\rm SN}/r_{\rm vir}$. For $N_{\rm SN}$, we have used equation \ref{eq:Nt_analytic}, which depends on mass. However, it suffices to use the initial halo mass because most of the supernova progenitors launching the wind must have formed in the early phases. Note that the virial radius scales as $\propto \xi^{1/3}$, so we have included a factor $\approx 1.25^{-1/3}$. Now, relating the pressure of the hot gas to a force on the shell, the maximum force from the supernovae is 

\begin{align}
F_{p, \rm max} & = \ 9.7 \times 10^{34} \mathrm{erg/cm} \left( \frac{f_{*}}{0.01} \right) \left( \frac{\kappa}{0.035} \right)^{-1} \left( \frac{m_{h}}{10^{8} M_{\odot}} \right)^{2/3} \left( \frac{f_{\rm cover}}{1} \right)  \\
    & \times 
    \begin{cases}
        \left( t_{\rm SF} / t_{\rm SN} \right)^{1.54} & \text{if } t_{\rm SF} \leq t_{\rm SN} \\
        (t_{\rm SF} - t_{\rm SN})/t_{\rm char} + 1 & \text{if } t_{\rm SF} > t_{\rm SN}.
    \end{cases}
\end{align}

\begin{figure}
    \centering
    \includegraphics[width=\linewidth]{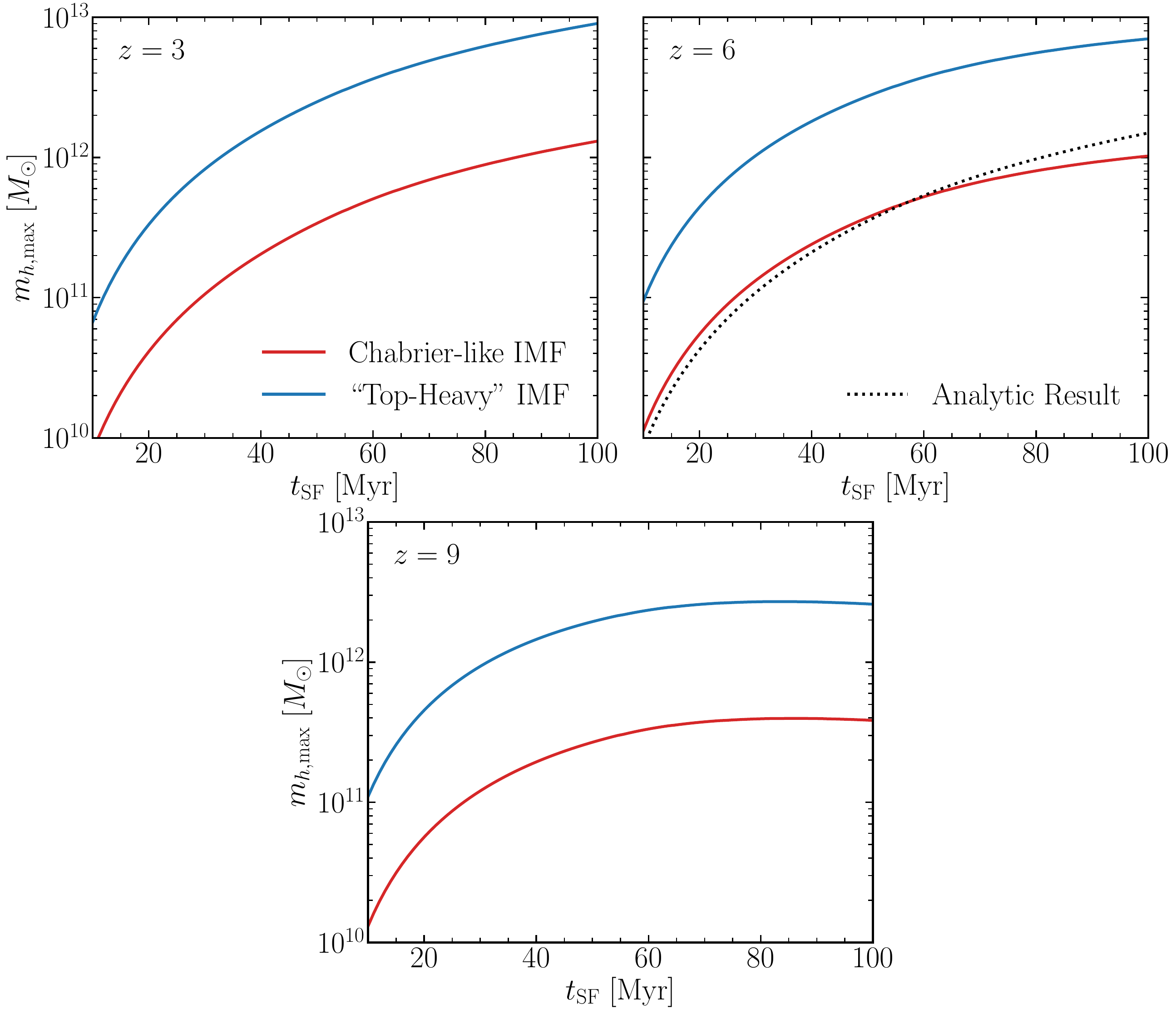}
    \caption{Maximum halo mass of mini-quenched galaxies when we simplify our model to include only the gravitational force and feedback by supernova. Stars are formed with an efficiency $f_{*} = 0.01$. The star-formation timescale, $t_{\rm SF}$, is on the $x$-axis. We plot the result of our analytic calculation (equation \ref{eq:mh_max}), which is performed for $z = 6$ galaxies forming stars in a Chabrier-like IMF. The analytic calculation does not implement halo growth in a robust way, so plotting this curve is meant to explore how including halo growth in $m_{h, \rm max}$ (solid lines) will differ from assuming constant halo mass.}
    \label{fig:mh_max_tmax}
\end{figure}

\subsection{The mini-quenching mass limit}  \label{sec:mh_max}

We expect there to be a maximum halo mass in our model because $F_{p} \propto m_{h}^{2/3}$ and $F_{g} \propto m_{h}^{4/3}$. To calculate this maximum halo mass, we begin by taking the ratio of the inward and outward forces:

\begin{align}
\frac{F_{p, \rm max}}{F_{g}} & = \ 270 \left( \frac{f_{*}}{0.01} \right) \left( \frac{m_{h}}{10^{8} M_{\odot}} \right)^{-2/3} \\
    & \times 
    \begin{cases}
        \left( t_{\rm SF} / t_{\rm SN} \right)^{1.54} & \text{if } t_{\rm SF} \leq t_{\rm SN}\\
        (t_{\rm SF} - t_{\rm SN})/t_{\rm char} + 1 & \text{if } t_{\rm SF} > t_{\rm SN}.
    \end{cases}
\end{align}
Quenching can no longer occur when the inward forces are greater than the outward forces on the shell which occurs at a maximum halo mass:

\begin{align}
m_{h, \rm max} & = 5 \times 10^{8} M_{\odot} \left( \frac{N_{\rm SN}}{1000} \right)^{3/5} \\
    & = 4.4 \times 10^{11} M_{\odot} \left( \frac{f_{*}}{0.01} \right)^{3/2} \\
    & \times 
    \begin{cases}
        \left( t_{\rm SF} / t_{\rm SN} \right)^{2.31}, & \text{if } t_{\rm SF} \leq t_{\rm SN} \\
        [(t_{\rm SF} - t_{\rm SN})/t_{\rm char} + 1]^{3/2},  & \text{if } t_{\rm SF} > t_{\rm SN}.
    \end{cases} \label{eq:mh_max}
\end{align}
One interesting result is that as long as the force of gravity is the dominant binding force ($\kappa \lesssim 0.4$ at $z = 6$), the maximum halo mass is not a function of the extent of the star forming region. This is not an obvious result because a more extended galaxy exists on average in a lower gravitational potential, so its gas should be easier to kick from the system. But at the same time, a larger extent of gas increases the volume that supernova must heat up to achieve a similar pressure. These two effects cancel one another in the parameter range we consider, and $m_{h, \rm max}$ does not depend on $\kappa$. 

A second (unsurprising) result is that this maximum halo mass value depends sensitively on the length of star formation. In short bursts of star formation, doubling the length of star formation allows for mini-quenching in halos five times larger!

In figure \ref{fig:mh_max_tmax}, we show how some key parameters affect $m_{h, \rm max}$ in this two-force model. Here, we no longer rely on scaling relations for the virial radius (equation \ref{eq:rvir}) or the number of supernovae (equation \ref{eq:Nt_analytic}), but solve these values numerically. In each panel of the figure, we display the maximum halo mass for $f_{*} = 0.01$ as a function of the star-formation timescale for several redshifts in which mini-quenched galaxies are observed. We plot the Chabrier-like IMF results as red lines and the representative ``top-heavy'' IMF as blue lines. While we would expect that a more top-heavy IMF would be more rare at $z = 3$ than, for example, $z = 9$, we plot these results in all panels for completeness. 

We compare the numerical estimate to the analytic result in the $z=6$ panel, where we show the latter with the black dotted line. Note the agreement at $t_{\rm SF} \sim 55$ Myr, which occurs because of our choice of the $\xi$ parameter; the estimate departs from the numerical results at larger times. 

To convert a maximum halo mass to a maximum stellar mass value, we assume that $m_{b} = (\Omega_{b}/\Omega_{m}) m_{h}$ and scale the baryon mass to a stellar mass by a time-averaged star formation efficiency $\tilde{f}_{*} = m_{*, \rm tot} / m_{b}$. With these conversions, we find 

\begin{align}
m_{*, \rm tot, max} & = 6.9 \times 10^{9} M_{\odot} \left( \frac{f_{*}}{0.01} \right)^{3/2} \left( \frac{\tilde{f}_{*}}{0.1} \right) \\
    & \times 
    \begin{cases}
        \left( t_{\rm SF} / t_{\rm SN} \right)^{2.31}, & \text{if } t_{\rm SF} \leq t_{\rm SN} \\
        [(t_{\rm SF} - t_{\rm SN})/t_{\rm char} + 1]^{3/2},  & \text{if } t_{\rm SF} > t_{\rm SN}.
    \end{cases} \label{eq:mstar_max}
\end{align}
This implies that, with isotropic accretion and typical feedback parameters, it is reasonable to expect mini-quenching in moderately large galaxies. For very efficient star formation, mini-quenching could extend to much higher masses.

\subsection{Inclusion of filaments in the approximation} \label{sec:mh_max_filaments}

Naively, we might not expect the inclusion of filaments to have significant implications on the maximum halo mass for quenching given the results in section \ref{sec:mh_max}. In that section, we found that the inward force is often dominated by gravity, while the outward force is dominated by supernova pressure. Both of these scale with $f_{\rm cover}$ and therefore cancel in our maximum mass equation. However, our approximations actually break down in the regime where $f_{\rm cover} \lesssim 0.1$. The inclusion of an $f_{\rm cover}$ factor in the force of gravity makes that force subdominant in many cases. When the accretion force is the dominant binding force, the maximum halo mass equation becomes

\begin{align}
m_{h, \rm max, fil} & = 1.9 \times 10^{10} M_{\odot} \left( \frac{f_{*}}{0.01} \right)^{3/2} \left( \frac{\kappa}{0.035} \right)^{-3/2} \left( \frac{f_{\rm cover}}{0.01} \right)^{3/2} \\
    & \times 
    \begin{cases}
        \left( t_{\rm SF} / t_{\rm SN} \right)^{2.31}, & \text{if } t_{\rm SF} \leq t_{\rm SN} \\
        [(t_{\rm SF} - t_{\rm SN})/t_{\rm char} + 1]^{3/2},  & \text{if } t_{\rm SF} > t_{\rm SN},
    \end{cases} \label{eq:mh_max_fil}
\end{align}
which is significantly smaller than the isotropic case. Interestingly, because the force of gravity is subdominant in this form of the model, the equation is now dependent on the initial radius of the shell. If the star-forming region of the galaxies are larger than our fiducial $\kappa = 0.035$ value by $10 \times$, then the maximum halo mass value decreases by $30 \times$. This result suggests that if filaments are present, then the star-forming region must be compact for mini-quenching to be effective.

\section{Reproducing mini-quenching observables with our model} \label{sec:quenching_timescale}

In this section, we expand our model to compare our results more rigorously with observations of mini-quenched galaxies and explore their implications for high-$z$ galaxy populations. For a fiducial model, we will assume a Chabrier-like IMF, $f_{*} = 0.01$, $f_{\rm cover} = 1$, $\kappa = 0.035$, and set the maximum star formation time to $t_{\rm SF} = 100$ Myr.

\begin{figure}
    \centering
    \includegraphics[width=\linewidth]{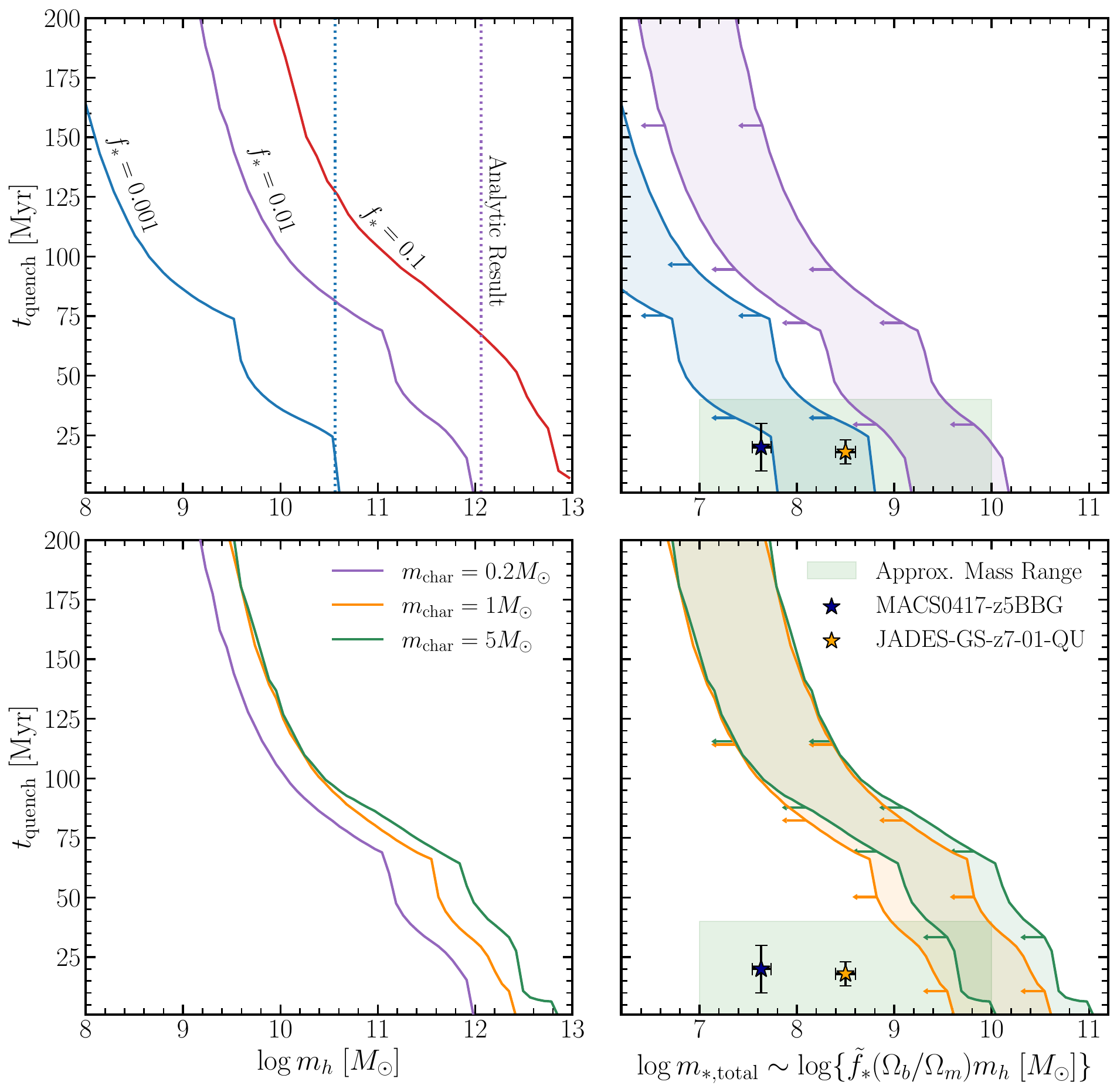}
    \caption{
    Quenching timescales at $z=6$ in our model. The left columns display $t_{\rm quench}$ as a function of halo mass, whereas the right columns convert to an approximate total stellar mass, $m_{*, \rm total}$ to compare with observations. 
    \textit{(Top row:)} Our fiducial model, with a Chabrier-like IMF and a maximum $t_{\rm SF} = 100$~Myr, for several choices of $f_{*}$. The vertical dotted lines show the result from equation \ref{eq:mh_max} for each $f_{*}$ value (with the same colors). In the top-right panel, we convert the $f_{*} = 0.001$ and $f_{*} = 0.01$ curves to stellar mass by assuming a net SFE in the range $\tilde{f}_{*} = [0.01, 0.1]$. The green square denotes the approximate mass range of observed mini-quenched galaxies (see section \ref{sec:quenching_timescale}), and the two stars correspond to the two example mini-quenched galaxies \cite{Strait2023, Looser2024} discussed in section \ref{sec:tquench_sfe}.
    \textit{(Bottom rows:)} Here we fix the SFE at $f_{*} = 0.01$ but change the IMF (as described in section \ref{sec:IMF}). We keep $\beta = 1.6$ and $\alpha = -2.35$ constant, but we increase $m_{\rm char}$ from the fiducial $m_{\rm char} = 0.2 M_{\odot}$ (purple line)  to $m_{\rm char} = 5 M_{\odot}$. In the bottom right panel, we only plot the increased $m_{\rm char}$ values for readability.}
    \label{fig:tquench}
\end{figure}

There are three observable results we wish to reproduce with our model: the stellar masses of mini-quenched galaxies ($m_{*, \rm total}$), the quenching timescales ($t_{\rm quench}$), and the amount of time mini-quenched galaxies were forming stars \textit{before} they became mini-quenched ($t_{\rm SF}$). Unfortunately, matching our model to stellar mass estimates is not a straightforward task, because there is not a single definition of ``mini-quenched.'' All of these quantities must also be extracted via SED fitting, so many have substantial uncertainties. Different authors apply the mini-quenching term to different populations \cite{Dome2024, Covelo-Paz2026, Yang2025}; we therefore consider a broad mass range of $m_{*, \rm total} = 10^{7}$--$10^{10} M_{\odot}$. The quenching timescale is similarly uncertain, if it is reported at all, though it is often found to be $t_{\rm quench} \sim 10$--$25$~Myr. The timescale is unlikely to be many times larger than this, or galaxies with longer quenching timescales would have been observed (though their rapid fading in the rest-UV does make this challenging). We therefore expect quenching for $\lesssim 40$ Myr to be a reasonable target. We illustrate this expected range of values with the light green regions in the right column of figure~\ref{fig:tquench}. 

We note that our model has several free parameters, while the observations of mini-quenched galaxies are so far limited to just a few examples. Thus our goal is not to develop a detailed set of constraints on the process but rather to examine the qualitative requirements set by mini-quenching in the context of our wind model.

\subsection{Recreating mini-quenching stellar masses and quenching timescales} \label{sec:tquench_sfe}

\begin{figure}
    \centering
    \includegraphics[width=0.6\linewidth]{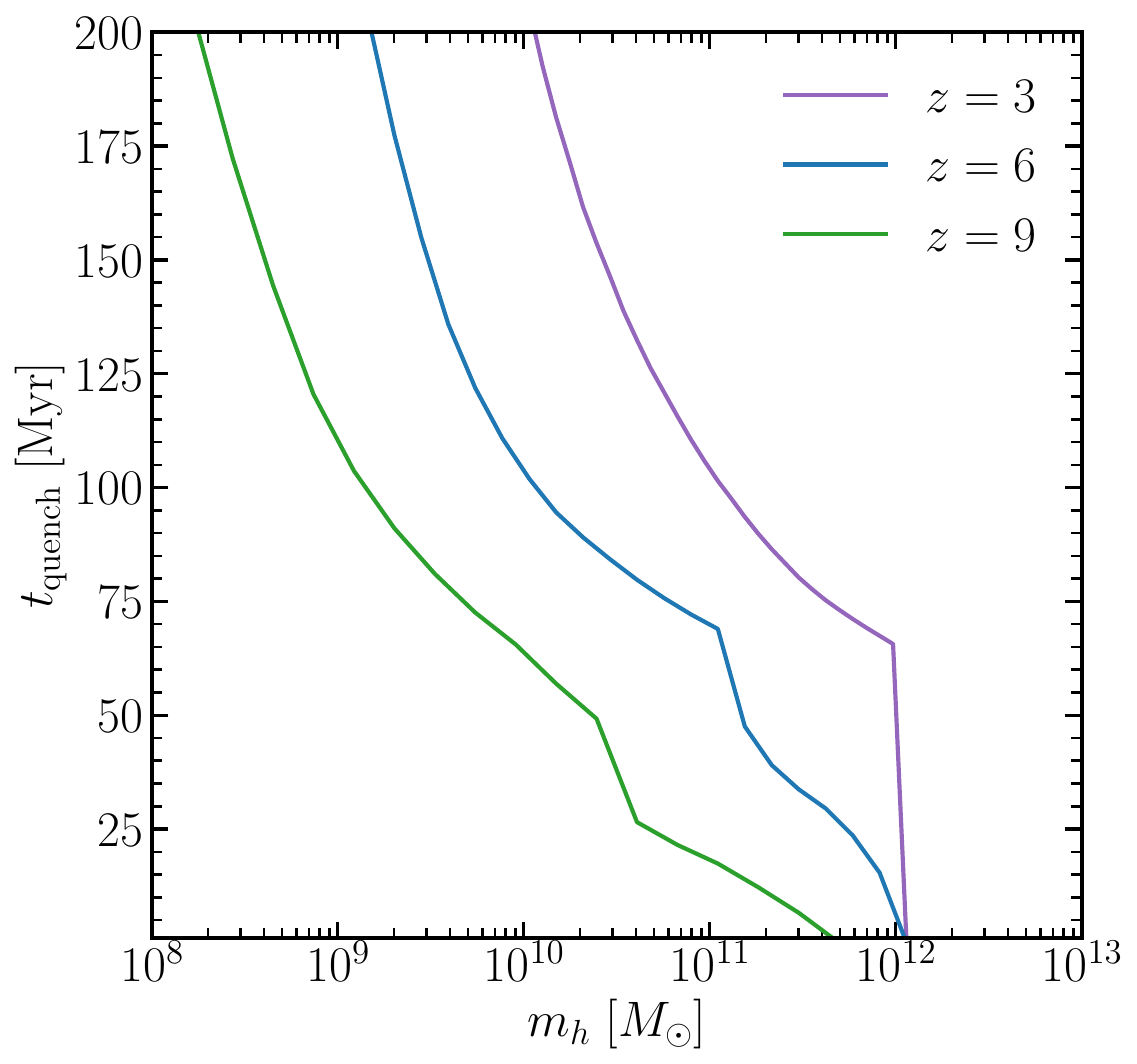}
    \caption{The effects of varying redshift in our model. To display this result, we plot the quenching timescales, $t_{\rm quench}$, as a function of the halo mass, $m_{h}$. In each case, galaxies form stars for $t_{\rm SF} = 100$ Myr, at $f_{*} = 0.001$, and with a Chabrier-like IMF. Briefly, we find that at later times, mini-quenching lasts for longer periods than at earlier times. This finding aligns with observations that mini-quenching lasts for $t_{\rm quench} \lesssim 50$ Myr at $z \sim 6$ (e.g., \cite{Gelli2023, Covelo-Paz2026}) and $t_{\rm quench} \lesssim 150$ Myr at $z \sim 3$ \cite{Merlin2025}.}
    \label{fig:tquench_diff_zs}
\end{figure}

In figure~\ref{fig:tquench}, we show how our model compares to observations of mini-quenched galaxies at $z=6$. In the top left panel, we plot the predicted quenching timescales for our model galaxies as a function of halo mass for several different SFEs. For reference, we plot the results of our equation \ref{eq:mh_max} prediction as vertical dotted lines for each SFE value. Then, to convert these results to stellar mass, we scale the halo masses by $m_{*, \rm total} \sim \tilde{f}_{*} (\Omega_{b}/\Omega_{m}) m_{h}$, where $\tilde{f}_{*}$ is the ratio of stellar mass to baryon mass. We plot each curve (except $f_{*} = 0.1$) with a range of $\tilde{f}_{*} = [0.01,0.1]$, which matches the range of typical values in models \cite{Diemer2015_colossus1, Furlanetto2017_minimalist, Prada2026} and observations \cite{Finkelstein2015}. 
In the top right panel, we find that our $f_{*} = 0.01$ curve can reproduce mini-quenching in nearly the entire range of masses and quenching timescales for our galaxies. In the same panel, we plot two example mini-quenched galaxies at $z \sim 6$. The first is JADES-GS-z7-01-QU \cite{Looser2024}, whose stellar mass and quenching timescale values we quote from their fitting with \textsc{bagpipes} \cite{Carnall2018}. (Note, however, that this galaxy is at $z = 7.3$; see figure \ref{fig:tquench_diff_zs} for the effect of redshift in our model). The second is MACS0417-z5BBG \cite{Strait2023}, where we follow their quoted uncertainties. Our model can easily halt star formation in these galaxies for the required intervals, even at their relatively high stellar masses, indicating that kinetic feedback can be more effective than thermal injection \cite{Gelli2024}. 

These results suggest that mini-quenching in small galaxies (including those highlighted in the figure) is relatively easy to achieve, as long as the star formation efficiency is not too small, although more massive systems that have been discussed are much more difficult to explain (the green box in the figure). However, we note that, so long as supernova feedback is dominant, the star formation efficiency and the wind efficiency ($\epsilon_{k}$) are degenerate. We assume here (optimistically) that $\epsilon_{k} = 1$, so that all of the supernova energy is available to drive the wind -- if not, the requirements for mini-quenching becomes more stringent, requiring reasonably large star formation efficiencies. We also emphasize that we have assumed isotropic accretion in this figure; below we will find that filaments can also make the requirements much more stringent. 

The ability to reproduce the stellar masses and quenching times also depends on the initial mass function in our model. In the bottom panels of figure \ref{fig:tquench}, we show how different IMFs change the maximum halo mass that can be quenched. When we increase the characteristic turnover in the IMF to $m_{\rm char} = 5 M_{\odot}$, the increased feedback efficiency produces mini-quenching in nearly the entire stellar mass range of mini-quenched galaxies (bottom right panel), as expected from section \ref{sec:mh_max}.  

\begin{figure}
    \centering
    \includegraphics[width=0.7\linewidth]{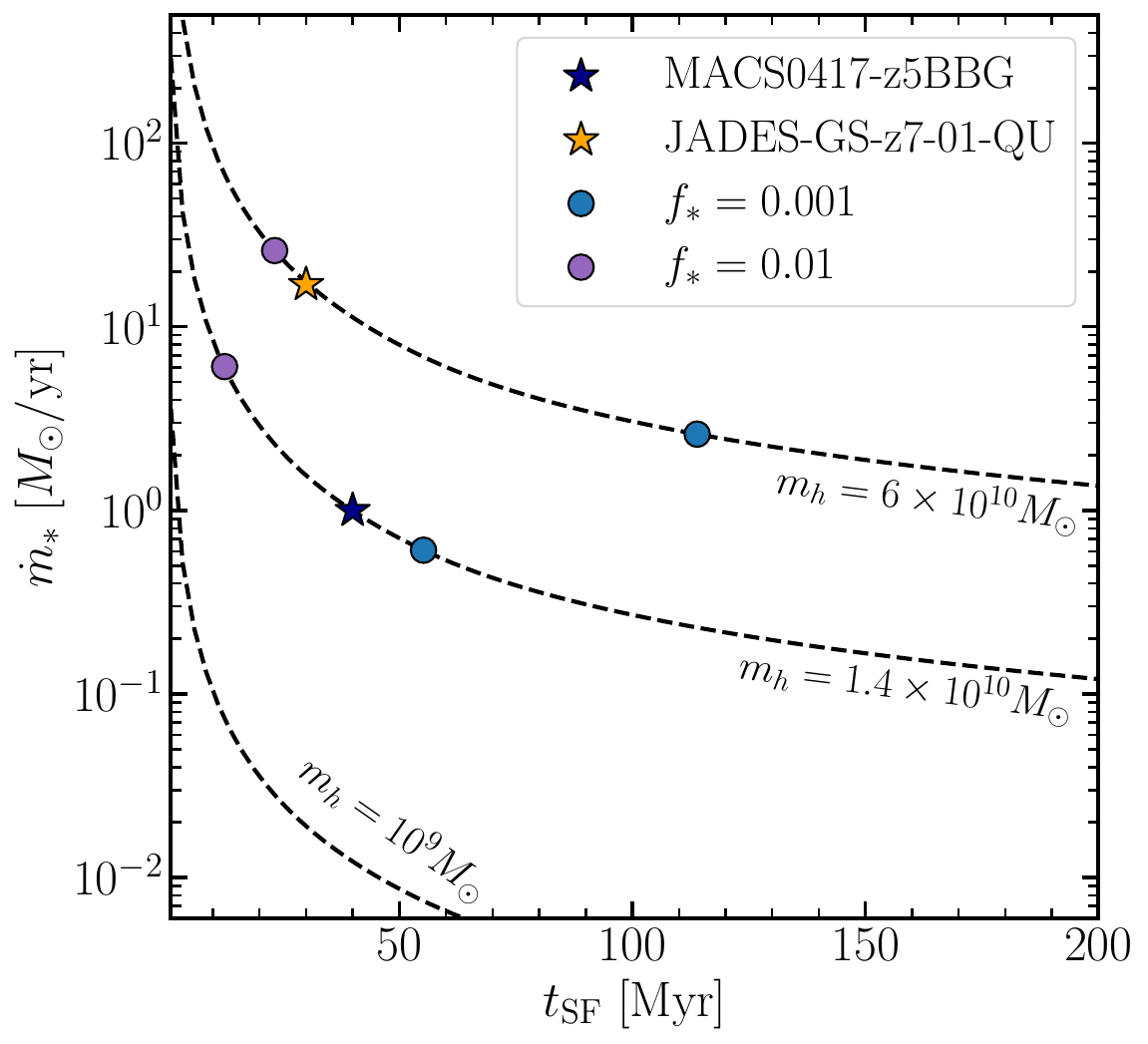}
    \caption{
    Requirements for the cumulative star formation to trigger mini-quenching, expressed in the plane of the duration of star formation ($x$-axis) and its rate ($y$-axis). 
    We compare our model's results to two galaxies, MACS0417-z5BBG \cite{Strait2023} ($z = 5.2$) and JADES-GS-z7-01-QU \cite{Looser2024} ($z = 7.3$), whose $\dot{m}_{*}$ and $t_{\rm max}$ have been estimated by \cite{Gelli2024} (stars). From our model, as black dashed lines, we plot the star-formation rate required to mini-quench galaxies with different halo masses and $t_{\rm SF}$ values. We plot two dashed curves with halo masses that correspond to the $\dot{m}_{*}$ and $t_{\rm SF}$ values of the observations. These halo masses should be seen as \textit{maximum} halo masses, and smaller halo masses with these star-formation rates would \textit{also} be quenched. Because $\dot{m}_{*}$ is a function of halo mass and $f_{*}$, we plot as dots the locations on the halo mass curves that correspond to $f_{*}=0.01$ and $f_{*} = 0.001$.}
    \label{fig:tmax_vs_mstar}
\end{figure}

We thus find that mini-quenching in moderately-massive galaxies can be achieved relatively easily. However, the timescales over which the quenching state can persist are very long for low-mass halos, which may conflict with observations. 

\subsection{Recreating star-formation timescales in mini-quenched galaxies}

The third and last observable our model must reproduce is the duration of star formation before the onset of mini-quenching, which we have labeled $t_{\rm SF}$. 

\cite{Gelli2024} estimated $t_{\rm SF}$
for two galaxies by fitting their spectral energy distributions, assuming the galaxy star formation histories are well described as top-hat functions before the quenching period. They found $t_{\rm SF} = 30$ Myr (JADES-GS-z7-01-QU, \cite{Strait2023}) and $t_{\rm SF} = 40$ Myr (MACS0417-z5BBG, \cite{Looser2024}), although we note that other evidence points towards being cautious about inferring star-formation histories from quenched galaxies (e.g., \cite{Suess2022, Haskell2024, Narayanan2024}). 

Our goal is to ensure our model can produce quenching quickly enough in galaxies with similar star-formation rate values as those observed. To do so, we return to the simplified model of section \ref{sec:approx}, but we use it to express the minimal requirements for mini-quenching in a different way: given a halo mass, we rearrange equation \ref{eq:mh_max} to solve for the combination of $f_\star$ and $t_{\rm SF}$ values that allow for mini-quenching. We then transform the SFE into the (observable) star formation rate, $\dot{m}_{*}$, using equation \ref{eq:mstar_dot}. We show the results of this approach in figure \ref{fig:tmax_vs_mstar}, with various halo masses as dashed curves. We emphasize that these curves show the \textit{minimum} star-formation rates that our model needs to have mini-quenching at those specific $t_{\rm SF}$ and $m_{h}$ values. Another way to phrase this is that for a given halo mass curve, our model can create mini-quenching in halos \textit{up to} that halo mass value. Mini-quenching can occur in halos with less mass for the associated $\dot{m}_{*}$ and $t_{\rm SF}$ values on the curve. 

To compare our results with observations, we plot the two observed galaxies as stars in figure \ref{fig:tmax_vs_mstar}. We see that our fiducial model can cause mini-quenching in each of these galaxies if their halo masses are \textit{below} the quoted values for the curves the observations lie on. To compare to the observations, however, we must convert these values to stellar masses. We find that the maximum halo mass calculated for MACS0417-z5BBG and JADES-GS-z7-01-QU match their quoted stellar mass values of $\tilde{f}_{*} \sim 0.02$ and $0.03$, respectively. Both of these values match model expectations \cite{Diemer2015_colossus1, Furlanetto2017_minimalist, Prada2026} and observations \cite{Finkelstein2015}, so we conclude that our model can produce mini-quenching on fast enough timescales to match observations.

The curves in figure \ref{fig:tmax_vs_mstar} are given for a constant halo mass, but for changing $\dot{m}_{*}$ values. Because the star-formation rate depends on both halo mass and $f_{*}$ (equation \ref{eq:mstar_dot}), movement along these curves corresponds to changing the star formation efficiency. Therefore, to guide the eye, we plot as dots two SFE values ($f_{*} = 0.01$ and $f_{*} = 0.001$). Both our example galaxies lie between these two values, suggesting that our model can create mini-quenching in galaxies with similar star-formation rates and $t_{\rm max}$ values as those observed, if the example galaxies are within that range of star-formation efficiencies. We note that although $f_\star$ is degenerate with the wind efficiency $\epsilon_k$ in terms of the feedback efficiency, its effects are probably more subtle on this plot because the latter does not affect the star formation rate itself. However, to a first order, $\epsilon_{k}$ still scales linearly with $f_{*}$, and therefore if $\epsilon_{k} = 0.1$ (rather than $\epsilon_{k} = 1$, as assumed here), the required $f_{*}$ values in the plot would be $10 \times$ larger.

We conclude that, with reasonable assumptions, mini-quenching following a star formation episode lasting $\sim 30$~Myr is relatively easy to achieve in our model. Thus, the key apparent challenge is understanding why the model predicts such long quenching timescales in small galaxies.

\subsection{Quenching timescales at different redshifts} \label{sec:diff_z}

We have so far focused on mini-quenching at $z \sim 6$, but it has also been observed at redshifts lower than $z \sim 6$. For example, 
\cite{Merlin2025} observed many low-mass  $z \sim 3$--$5$ galaxies that have been dormant for $t_{\rm quench} \lesssim 150$ Myr. There should also be mini-quenching at \textit{higher} redshifts, but these galaxies are difficult to observe because the key observational quantity to confirm low star-formation rates (the H$\alpha$ line) has redshifted out of detectability by $z \sim 7.4$ for many instruments \cite{Covelo-Paz2026}. 

Here, we explore how mini-quenching changes as a function of redshift. We must turn to our full numerical model to test the redshift dependence because assumptions in our simple analytic result break down at earlier times. For example, the accretion rate scales as $\dot{m}_{h} \propto (1+z)^{5/2}$, and Compton cooling, whose rate scales as $t_{\rm comp} \propto (1+z)^{-4}$, becomes important at earlier times (see appendix \ref{sec:compton_cooling_approx}).

Figure \ref{fig:tquench_diff_zs} shows the quenching timescales of our model at a few different redshifts. Here the complex effects of redshift are apparent: the maximum mass of mini-quenched galaxies changes slightly from $z = 6$ to $z = 3$, though it changes drastically from $z = 6$ to $z = 9$. Lower redshifts produce longer quenching timescales as well. This result aligns with the trend that at $z \sim 6$, quenching timescales appear to be $\lesssim 40$ Myr, while at $z \sim 3$, these timescales have increased to $\lesssim 150$ Myr. While it is true that this lower-redshift sample could be contaminated with sources that are not mini-quenched but will be completely quenched (which can be the case even for low-mass early galaxies; see \cite{Baker2025_environment}), it is reassuring that the timescale that galaxies are mini-quenched at $z = 3$ is longer than the mini-quenching at $z = 6$. 

\begin{figure}
    \centering
    \includegraphics[width=\linewidth]{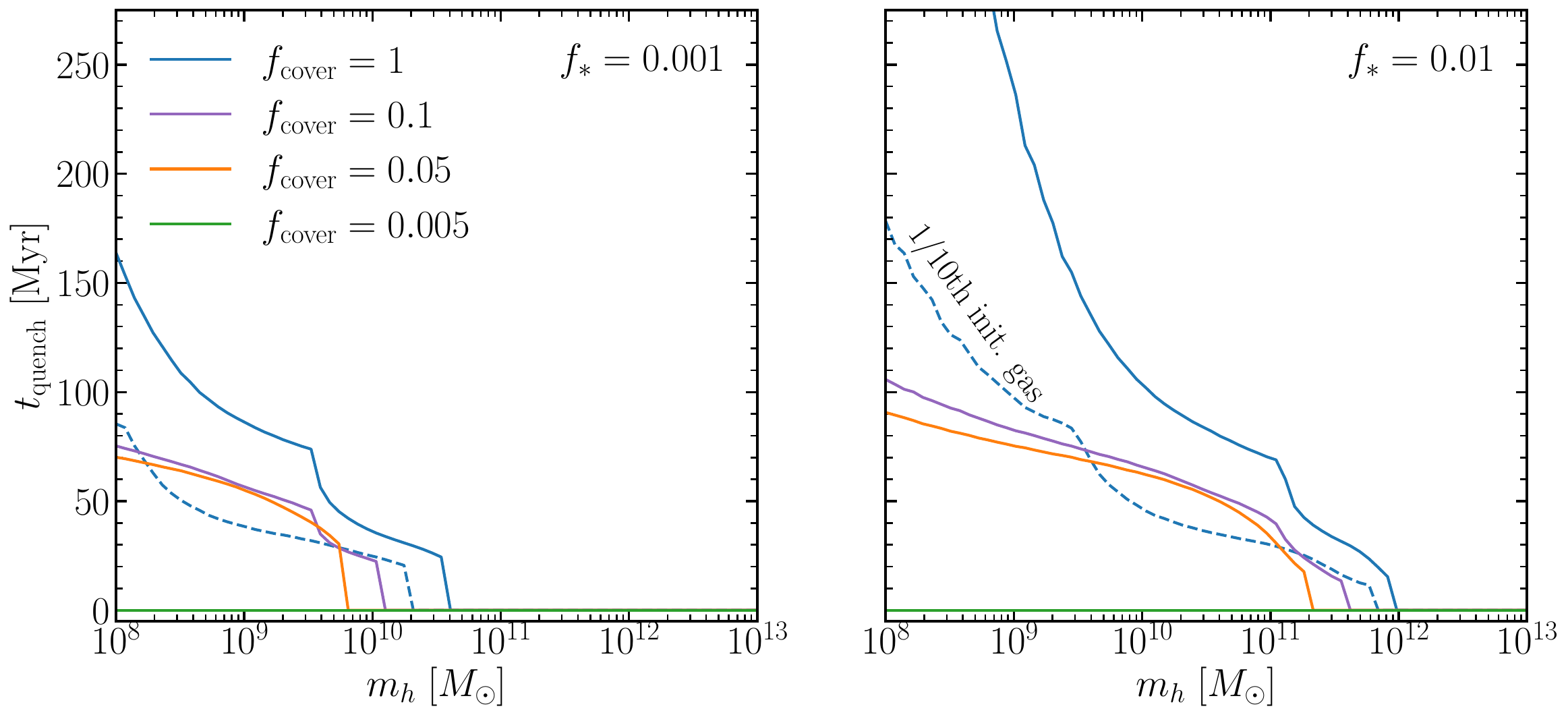}
    \caption{
    The impact of filamentary accretion on mini-quenching. We plot these results for two values of $f_{*}$ (left and right panels). For each halo mass, $m_{h}$, we plot the quenching timescale, $t_{\rm quench}$, on the $y$-axis. We assume a Chabrier-like IMF with $t_{\rm SF} = 100$ Myr in all cases and assume $z=6$. Each colored curve represents a different covering fraction, $f_{\rm cover}$ (see section \ref{sec:filaments}). The blue curve shows the result of our fiducial model (which implicitly assumes $f_{\rm cover} = 1$). We decrease the covering fraction down to $f_{\rm cover} = 0.005$, which is the approximate covering fraction using the filament model from \cite{Mandelker2018}. For such narrow filaments, galaxies do not undergo mini-quenching in our model ($t_{\rm quench}=0$ across all the halo masses). Further, we plot as dashed blue lines the result of initializing our $f_{\rm cover}=1$ model with $10 \%$ of the gas mass as the rest of our models, also decreasing the quenching timescale.}
    \label{fig:tquench_diff_fcovers}
\end{figure}

\subsection{The effects of filamentary accretion} \label{sec:filaments}

A result of our model that should not be overlooked is that it predicts very long quenching timescales for many halo masses (see figure \ref{fig:tquench}). Observations do allow $t_{\rm quench} \gtrsim 100$ Myr at $z = 6$, given that mini-quenched galaxies with longer timescales are likely more difficult to identify, so we may eventually observe some of them. However, the apparent discrepancy at low masses is large, so in this section, we explore how the inclusion of filaments ($f_{\rm cover} < 1$) affects the maximum quenching timescale.

Figure~\ref{fig:tquench_diff_fcovers} shows the impact of the size of filaments on mini-quenching timescales. In this figure, we run our model for two values of $f_{*}$ (left and right panels) and with filaments of different sizes (parameterized by changing $f_{\rm cover}$). Recall that the accretion rate is held constant as we change $f_{\rm cover}$, so the primary effect of accretion is to increase the ``momentum-loading'' term in the wind propagation equation.  The blue curves represent the model with isotropic accretion, or $f_{\rm cover} = 1$. When we decrease $f_{\rm cover}$, both the maximum halo mass and the quenching timescale drop dramatically. If we assume $f_{\rm cover} = 0.005$ (green line), as implied for all $z = 6$ observed galaxies by the model from \cite{Mandelker2018}, mini-quenching becomes impossible. These results show the drastic effect filamentary accretion by gas has on mini-quenching in our model. 

While filamentary accretion significantly decreases the maximum halo mass that is quenched in our model, it also decreases the quenching timescale. If $f_{\rm cover} = 0.1$, then the quenching timescale decreases from $t_{\rm quench} > 100$ Myr to $t_{\rm quench} \lesssim 50$ Myr in many cases. This change improves the agreement with observed systems because it increases the importance of accreting gas on the motion of the shell, which does not decrease with radius like gravity does.

Interestingly, we find that in the left panel of figure \ref{fig:tquench_diff_fcovers}, a lower covering fraction can create a slightly larger quenching timescale for a range of halo masses. This effect is the result of the complicated interplay between the time spent forming stars and the quenching timescale. In a scenario with a smaller covering fraction, the galaxies spend more time forming stars before the galaxy becomes quenched, and in this range of halo masses, these extra stars result in longer quenched periods.

Thus, we find that narrow filamentary accretion has a dramatic effect on mini-quenching. Filaments as narrow as suggested by some models make it essentially impossible, because the incoming gas has such a high ram pressure. Mini-quenching can thus only occur if accretion at high redshifts occurs over extended filaments. That said, such filaments can decrease the timescale of quenching at low halo masses, alleviating one of the challenges of the model.

\subsection{Sensitivity to initial conditions} 

We also plot in figure~\ref{fig:tquench_diff_fcovers} as dashed blue curves the result of a model with $f_{\rm cover} = 1$ but initialized with $10 \%$ of the initial gas mass. Such a setup might occur if a galaxy removed most of its gas in the previous quenching cycle. In this case, the quenching timescale decreases drastically. Therefore, another potential solution for the large quenching timescales at low halo masses is that these galaxies have already evacuated most of their gas in previous cycles.
But such an occurrence also serves to decrease the star-formation rates of these galaxies, potentially causing disagreement in other observational quantities. For example, figure \ref{fig:tmax_vs_mstar} displays the initial conditions that produce star-formation rates and $t_{\rm SF}$ which agree with observations. If we tweak the calculation so that the gas mass is $10 \%$ of what we assumed in the figure, keeping all the other parameters constant, the required $t_{\rm SF}$ becomes $\sim 2$--$3 \times$ larger, disrupting the agreement with our model and the observations. Of course, we could compensate for this problem by adjusting other model parameters, such as by increasing the star-formation efficiency by $10 \times$ or the halo mass by $\sim 4 \times$. These changes restore the agreement in figure \ref{fig:tmax_vs_mstar}, although these secondary changes have cascading effects (for example, by increasing the quenching timescales as well). Therefore, while initializing a galaxy with less interstellar gas is a solution to our too large quenching timescales at low masses, it does not appear to be a robust solution.

\begin{figure}
    \centering
    \includegraphics[width=\linewidth]{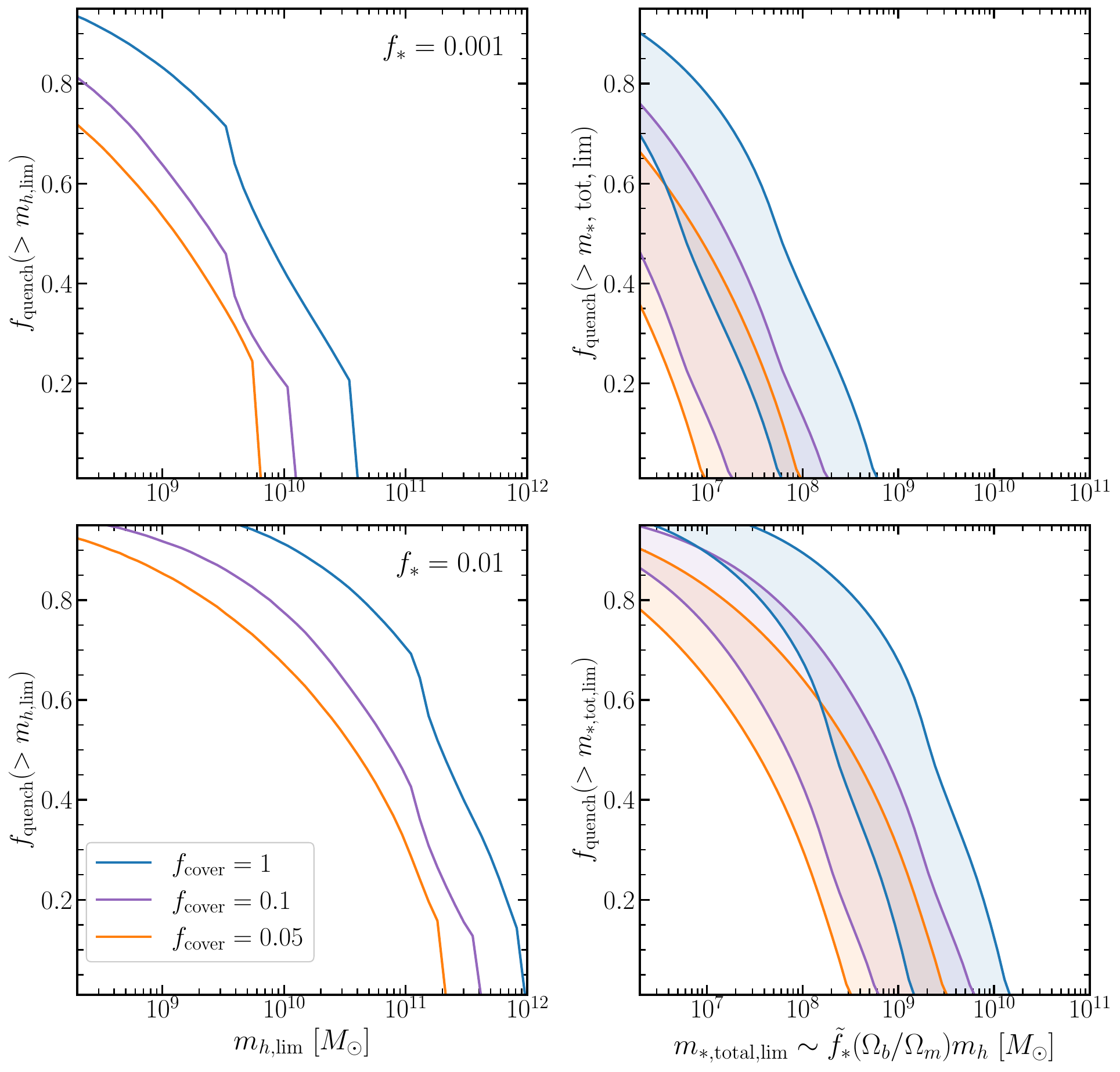}
    \caption{
    The fraction of galaxies undergoing mini-quenching as a function of minimum observed halo and stellar mass, assuming all galaxies above that mass threshold are observed. This fraction is the duty cycle of the quenching phase, weighted by the halo mass function (see equation~\ref{eq:f_quench}). We show it as a function of SFE and the covering fraction of the accreting filaments.  Briefly, the quenching fractions are high in our model if accretion is isotropic, though filamentary accretion allows for smaller fractions -- filaments appear to be required at $z \sim 6$ if only a small fraction of galaxies are mini-quenched. All curves take $z=6$ and assume a Chabrier-like IMF.}
    \label{fig:fquench_fcovers}
\end{figure}

\subsection{Estimating the fraction of galaxies in mini-quenched phases} \label{sec:fquench}

So far, we have focused on mini-quenching in individual galaxies, but an interesting question is how common these systems are. A complete census of mini-quenched galaxies across the observed samples is currently very difficult, because of the challenges in identifying these systems and interpreting their spectra (though see \cite{Yang2025, Khullar2026}). On the theory side, simulations can be used, though they fail to reproduce all the characteristics of these galaxies in, e.g., the shape of quenched SEDs \cite{Gelli2023, Dome2024}. 

The fraction of time a galaxy is quenched (or its duty cycle) is $f_{\mathrm{quench}, i} \approx t_{\rm quench} / (t_{\rm quench} + t_{\rm SF})$. This quantity can be folded together with the halo mass function to find the total fraction of galaxies that will be quenched at any given time:

\begin{equation}
    f_{\rm quench} = \frac{\int_{\log m_{h, \rm min}}^{\infty} f_{\mathrm{quench}, i}(m_{h}) n_h(m_{h}) d\log m_{h}}{\int_{\log m_{h, \rm min}}^{\infty} \Phi(m_{h}) d\log m_{h}},
    \label{eq:f_quench}
\end{equation}
where $n_h$ is the halo mass function in units of number density per logarithmic interval of mass and $m_{h, \rm lim}$ is the minimum halo mass observable by a survey. To compute this quantity, we use the \cite{Watson2013_HMF} halo mass function with the friends-of-friends halo-finding method \cite{Davis1985_fof}, though these choices will not significantly alter our results.

We use our full model to estimate this fraction in figure \ref{fig:fquench_fcovers} (with a Chabrier-like IMF at $z = 6$). The first row assumes $f_{*} = 0.001$, while the second row takes $f_{*} = 0.01$. In each panel, we show three different cases for the accretion geometry, including isotropic accretion and filaments with different covering fractions. 
The second column takes the halo mass results in the first column and converts them to an approximate total stellar-mass range, assuming $\tilde{f}_{*} = [0.01,0.1]$. We find that a small $f_{\rm quench}$ among low-mass galaxies only occurs if filaments are very narrow ($f_{\rm cover} \ll 1$). 

Direct, observational tests for filaments at high redshift are impossible at the moment, though indirect tests might be possible. For example, our model predicts that galaxies accreting isotropically will be mini-quenched for much longer periods than galaxies accreting by gas filaments. Therefore, our model predicts that \textit{field} galaxies may be mini-quenched for longer periods than galaxies in the Cosmic Web, if one makes the reasonable assumption that field galaxies accrete more isotropically than galaxies embedded in the Cosmic Web (which accrete by filaments). Simulations could test this prediction as well, though the simulation would have to span cosmological scales to securely estimate $f_{\rm quench}$, while still having a sufficiently high resolution  to resolve outflows. Certainly, future work is required, and within our current capabilities.

\section{Discussion} \label{sec:discussion}

Our model is by construction quite simple. Here we consider a few well-motivated extensions that could help explain other observable features of the interactions of winds and accretion in mini-quenching. 

\subsection{Instabilities in the expanding shell} 

So far, we have assumed that mini-quenching continues until the feedback-driven shell stalls and falls back. However, it is very likely that these shells are actually part of complex multiphase media which impact the equally complex and multiphase CGM surrounding their host galaxies, including accreting streams. The complex flows can trigger instabilities in the wind, potentially allowing some fraction of the accreting material to penetrate the shell and resume star formation. 

Here, we estimate the timescale over which one such instability may develop to be a limiting factor of the duration of mini-quenching episodes. We suppose the shell and the accreting gas experience Rayleigh-Taylor instabilities, as the low-density wind material collides with the surrounding CGM inside the halo gravitational potential well. This instability develops over a timescale $\tau_{\rm RT} \equiv 2\pi/\omega$ \cite{GalaxyFormation&Evolution2010}, where $\omega =[g k (\delta/(2 + \delta)]^{1/2}$ is the linear growth rate of the instability and $\delta = 1 - \rho_{\rm sh} / \rho_{\rm acc}$ is the overdensity of the shell relative to the accreting material. In this case, $g$ is the gravitational acceleration at the boundary of the two mediums and $k = 2 \pi / r$ is the wave number of the growing perturbation. 

With $r \approx R_{0} = \kappa_{\rm in} r_{\rm vir}$ being the initial radius of our shell and the gravitational acceleration being $g = 2Gm_{h}/\kappa r_{\rm vir}^{2}$, the timescale for instabilities to grow is

\begin{equation}
    \tau_{\rm RT} = \frac{2 \pi}{\omega} = 32 \ \mathrm{Myr} \left( \frac{\delta}{2 + \delta} \right)^{-1/2}\left( \frac{\kappa_{\rm in}}{0.07} \right) \left( \frac{1 + z}{7} \right)^{-3/2},
\end{equation}
where we have chosen $\kappa_{\rm in}$ to be twice the fiducial starting shell value.
If the accreting material is a few times denser than the shell (as is the case for strongly shocked material), $\delta \sim 3/4$ and $\tau_{\rm RT} \sim 60$ Myr. At $z = 3$, $\tau_{\rm RT} \sim 140$ Myr. This is only a very rough estimate for the instability timescale (and others, such as the Kelvin-Helmholtz instability, may also be relevant). Nevertheless, it is intriguing that these instability timescales roughly match estimates of the quenching timescales at both $z \sim 6$ and $z \sim 3$ (e.g., \cite{Covelo-Paz2026, Merlin2025}), suggesting that instabilities can help limit the extremely long quenching periods predicted by our model for small galaxies. 

If we perform a similar process as section \ref{sec:fquench}, a ceiling of $t_{\rm quench} = 30$ Myr decreases the quenching fraction by only $\sim 20 \%$ up until $m_{h, \rm lim} \sim 10^{11} M_{\odot}$. At higher masses, the average quenching timescale of galaxies is already below 30 Myr, and the quenching fraction is unaffected. This result suggests that instabilities are an important part of mini-quenching, and understanding the dynamics of outflows is important in connecting the incidence of mini-quenching to theory models.

\subsection{Can AGNs increase mini-quenching?}

To this point, we have only considered stellar feedback, but an accretion episode could also trigger an AGN. In this section, we consider how AGN feedback may contribute to mini-quenching. This question is particularly interesting if the observed early galaxies accrete through thin filaments, for which supernova winds are far less effective. To explore this possibility, we use a simple analytic estimate similar to that of section~\ref{sec:approx}. We begin by assuming that the black hole is accreting at the Eddington luminosity, $L_{\rm edd}$, and that a fraction of the luminosity produced by the black hole provides feedback, $\epsilon_{\rm F}$. In this case, the feedback force produced by the black hole is $F_{\rm BH} = f_{\rm cover} \epsilon_{\rm F} L_{\rm edd} / c$.\footnote{Here we have assumed that the black hole feedback is itself isotropic. In most models of AGN feedback, winds and jets are not isotropic, but in that case they would only help with mini-quenching if the feedback and accretion flows were aligned. We ignore this possibility for simplicity.} If the black hole is the only source of feedback and the force of accretion is much greater than the force of gravity (section \ref{sec:mh_max_filaments}), the galaxy is quenched if

\begin{equation}
    f_{\rm cover} \epsilon_{\rm F} \frac{L_{\rm edd}}{c} - \frac{dm_{\rm sh}}{dt} v_{\rm vir} > 0.
\end{equation}
The Eddington luminosity is $L_{\rm edd} = 4 \pi G c m_{p} m_{\rm BH} / \sigma_{\rm T} \sim 1.3 \times 10^{46} \mathrm{[erg/s]} (m_{\rm BH}/10^{8} M_{\odot})$, where $m_{p}$ is the mass of the proton and $\sigma_{\rm T}$ is the Thomson cross section. If the halo mass of the galaxy relates to the total stellar mass by $m_{h} = m_{*, \rm tot}(\Omega_{m}/\Omega_{b}) / \tilde{f}_{*}$, the equation above can be used to estimate the largest stellar mass a galaxy can have to be mini-quenched:

\begin{equation}
    m_{*, \rm tot, AGN} \approx 2.5 \times 10^7 \,  M_{\odot} \ f_{\rm cover}^{3} \epsilon_{\rm F}^3 \left( \frac{m_{\rm BH} / m_{*, \rm tot}}{0.01} \right)^{3} \left( \frac{\tilde{f}_{*}}{0.1} \right)^{4}.
    \label{eq:mtot_AGN} 
\end{equation}
Here we chosen a black-hole mass to stellar mass ratio of $m_{\rm BH}/m_{*, \rm tot} = 0.01$, which agrees with observations of early black holes \cite{Harikane2023_AGN, Pacucci2023_Mbh-Mstar, Juodvbalis2025_AGN}, but is large compared to local galaxies. Comparison to figure~\ref{fig:tquench} shows that AGNs are somewhat less efficient than stellar feedback, which is not surprising on energetic grounds -- while black hole accretion is about an order of magnitude more efficient than stellar fusion, the black holes are still much smaller than the host galaxies in our fiducial scaling.  

The steep scaling with $f_{\rm cover}$ implies that the additional AGN feedback still would not be enough to trigger mini-quenching if $f_{\rm cover} \sim 0.005$, as suggested by the analytic model of \cite{Mandelker2018}. Our model therefore suggests that mini-quenched galaxies have relatively diffuse accretion channels. 

\subsection{Escaping winds and metal enrichment} 

One of the most widely-recognized implications of stellar feedback is large-scale chemical enrichment of the early Universe. Estimates of wind enrichment using feedback models similar to our own -- with the important exception of ignoring the accretion flows and hence allowing continuous star formation -- have shown that a substantial fraction of the IGM can be enriched by metals in this way (e.g., \cite{Ferrara2000, Madau2001, Furlanetto2003_winds, Yamaguchi2023}). In contrast, in our model only the smallest halos are able to liberate the gas from the halo.

As noted, the key difference is in our treatment of accretion flows. Once mini-quenching occurs, and the initial wave of supernovae cease, the inward momentum added by accretion will inevitably stall the shell in all but the smallest systems. To estimate the maximum halo mass whose wind can escape the system (in isotropic accretion),
the binding energy required to move a shell to the virial radius can be  approximated as $E_{b} = E_{g} + E_{\rm acc}$. We have seen that gravity dominates the inward force for $r \lesssim 0.32 r_{\rm vir}$, and accretion dominates outside of that radius. Thus, we can estimate the energy required as the work to overcome the gravitational force to $0.32 r_{\rm vir}$ and the work to overcome the accretion force at larger radii. The energy pushing against the shell is at maximum the total energy from supernova, $E_{\rm input} = \epsilon_{\rm SN} N_{\rm SN}\times 10^{51}$erg (see appendix \ref{sec:N_analytic} for an analytic calculation of $N_{\rm SN}$). If we set the binding energy equal to the input energy, we find the maximum halo mass that gas can escape an isotropically-accreting galaxy is

\begin{equation}
    m_{h} \lesssim 1 \times 10^{11} M_{\odot} \left( \frac{f_{*}}{0.01} \right)^{3/2},
\end{equation}
assuming a $z = 6$ galaxy forming stars for $t_{\rm SF} = 100$~Myr with a Chabrier-like IMF, and $\epsilon_{\rm SN} = 1$. Therefore, observed mini-quenched systems can remove gas from a galaxy, even in the presence of an accretion force. However, if the accreting gas has the same velocity outside the virial radius as it does inside it, the gas will eventually recycle back into the galaxy because the accretion force will continue to push on the outflowing gas outside the halo.

In the case of filaments, much of the gas will not encounter an accretion force because it is not incident on the accreting filaments. The outflowing gas will be subject only to the gravitational force and thus have an easier time leaving the galaxy. The maximum halo mass where the gas can escape the galaxy becomes $m_{h} \lesssim 1.5 \times 10^{11} M_{\odot} (f_{*}/0.01)^{3/2}$, which is not a sizable difference. Further, this maximum halo mass resembles the values found in \cite{Furlanetto2003_winds}, which used a similar wind model, albeit without including the accretion force we have highlighted here.

\subsection{Starbursts following a mini-quenching episode} \label{sec:reaccretion}

We now consider another consequence of mini-quenching, relevant for any case in which the shell gas remains bound to the system. When a galaxy is mini-quenched, feedback from the galaxy (of any form) depletes the gas from the galaxy and prevents any further gas from accreting for the remainder of its quenching period. Afterwards, the galaxy begins to accrete again. But what happens to the gas that has been prevented from accreting onto the galaxy?

In the previous section, we showed that when accretion is included, mini-quenching can only \emph{temporarily} prevent gas from falling back onto a galaxy, except at the smallest masses. Using the galaxy growth model in section \ref{sec:growth_model}, the amount of gas that will accrete over a quenching period, $t_{\rm quench}$, is

\begin{equation}
	\Delta m_{\rm acc} = m_{g, i} (1 - \exp [ A t_{\rm quench} (1 + z)^{5/2} ] ),
\end{equation}
where $m_{g, i}$ is the initial gas mass in the galaxy before mini-quenching. Thus mini-quenching for $\sim 50$~Myr will temporarily prevent a gas mass $\gtrsim 0.2 (\Omega_b/\Omega_m) m_h$ from accreting (or larger at $z \gg 6$)  -- which will then fall back quickly once the mini-quenching period ends, likely triggering another bout of rapid star formation. 

\cite{Sun2026} found that early galaxies can form stars (at lower rates) out to $\sim 0.4 r_{vir}$. While their calculation is contextualized around star formation in a turbulent circumgalactic medium at early times, their results might apply to this scenario as well. Stars might form in the gas during infall, and there could be a large amount of star formation at once. However, a more sophisticated analysis will be required to determine the outcome of this large gas mass that will accrete post-quenching. 

\section{Conclusions} \label{sec:conclusion}

We have presented a model to explore how star formation can create mini-quenching in high-$z$ galaxies. Below, we lay out our key conclusions and their implications for mini-quenching.

We found that our model is able to reproduce key observables in mini-quenched galaxies with feedback from star formation. Galaxies in our model produce stellar feedback through a combination of dust-driven winds, radiation pressure by Ly$\alpha$ photons, and a pressure force created by supernova, with the last typically dominating. If these feedback forces overcome the binding forces of gravity and accretion (included explicitly in our model for the first time), the galaxy is quenched and the gas in the galaxy is pushed into a thin shell outside the star-forming region. The shell is expanded by feedback but eventually falls onto the galaxy again once star formation ceases and the feedback force fades. Using this model, we find:

\begin{enumerate}
    \item Adopting plausible galaxy-formation parameter values (e.g., $f_{*} = 0.01)$, supernovae provide enough feedback to remove gas from galaxies with similar stellar masses to observed mini-quenched galaxies at Cosmic Dawn -- up to $m_{*, \rm tot} \approx 10^{10} M_{\odot}$ (figure \ref{fig:tquench}), if accretion is isotropic rather than filamentary. This mass scale is sensitive to both the adopted star-formation efficiency (e.g., increasing $f_{*}$ by a factor of ten increases the mass scale by a factor of $\sim 15$) and the details of star formation (e.g., a top-heavy IMF can increase the threshold by up to $\sim 8 \times$).
    \item Stellar feedback from a galaxy lasts long enough to reproduce the $\lesssim 50$ Myr quenching timescales of mini-quenched galaxies at $z \sim 6$ (figure \ref{fig:tquench}). In fact, feedback in our model causes quenching for $t_{\rm quench} \gtrsim 100$ Myr with certain parameter choices, which is \textit{too} long.
    \item One possibility to reduce the quenching timescale is for the galaxy to accrete by filaments (figure \ref{fig:tquench_diff_fcovers}). However, this is a balancing act, as filaments must be much larger than expected sizes from analytic models, or else the galaxy would not be able to prevent further accretion of cold gas and mini-quench the galaxy. 
    A second solution is to appeal to instabilities in the expanding shell, which could prevent galaxies from being quenched for longer than $\sim 30$--$100$~Myr. Such a solution also predicts that galaxies cannot be mini-quenched for long periods -- in agreement with current (though biased) observations.
    \item Supernovae provide energy at fast enough rates to mini-quench $z = 6$ galaxies with star-formation periods equal to or less than observed values (figure \ref{fig:tmax_vs_mstar}). If the observed star-formation timescales are correct, our model can mini-quench galaxies with a stellar mass to baryon mass ratio of $\sim 0.02$ -- well within the range of expected and observed values. 
    \item Our model predicts that if galaxies accrete with thin filaments, then the fraction of galaxies that are mini-quenched is small. If a population of galaxies accretes isotropically (e.g., as field galaxies might), then their quenching timescales are long.
\end{enumerate}

Future observations will be key to understand mini-quenched galaxies. Currently, our knowledge of the process is limited by the sparsity of observed systems. Upcoming surveys will no doubt reveal galaxies at all stages of quenching, helping us statistically measure the timescales associated with quenching, and therefore constraining the underlying galaxy formation parameters. Further, simulations capable of resolving the complicated dynamics of turbulent, interstellar gas at the edge of a galaxy, as well as the Cosmic Web as a whole, are forthcoming. Such simulations will illuminate the outer reaches of a galaxy, and advance our understanding of how outflowing gas interacts with accreting gas. In the future, we expect such improvements in understanding mini-quenching to provide a valuable window into early star formation.

\appendix

\section{Analytic calculation of the number of supernova in our model} \label{sec:N_analytic}

In this section, we calculate the number of supernovae that will occur after some time $t$ of star formation. We assume there has been no previous star formation, so that only the stars forming at $t \geq  0$ produce supernovae. In this calculation, we assume a Chabrier-like IMF (see section \ref{sec:IMF}) and a galaxy at $z = 6$. 

First, we will assume that stars have only been forming for $t < t_{\rm SN}$, where $t_{\rm SN} = 55$ Myr (the lifetime of the smallest star to go supernova). At times longer than this, all the supernova will have gone off for the stars formed at $t = 0$ Myr. This adds complexity to the calculation, which we will expand upon later in this section.

We can calculate the rate of supernova occurring for a population of age $\Delta t$ using equation~\ref{eq:dNsn_dt}:

\begin{equation}
    \dot{N} (\Delta t) = 30 \ \mathrm{Myr}^{-2} \left( \frac{f_{*}}{0.01} \right) \left( \frac{m_{h}}{10^{8} M_{\odot}} \right) \int_{0}^{\Delta t} dt' \left( \frac{t'}{\rm Myr} \right)^{-0.46},
\end{equation}
where we have assumed a constant halo mass for any given $t_{\rm max}$ (though we include a factor to account for halo growth in equation \ref{eq:mh_max}). With this simplified equation, we compute a scaling relation for the rate of supernovae as a function of time and other parameters:

\begin{equation}
    \dot{N}(\Delta t) = 56 \ \mathrm{Myr}^{-1} \left( \frac{\Delta t}{\rm Myr} \right)^{0.54} \left( \frac{f_{*}}{0.01} \right) \left( \frac{m_{h}}{10^{8} M_{\odot}} \right). \label{eq:Ndot_analytic}
\end{equation}
The total number of supernova to come after a time $t$ is achieved by integrating equation \ref{eq:Ndot_analytic} such that 

\begin{align}
    N(t) = & \int_{0}^{t} \dot{N}(t-t') dt' \\
      = & \  36 \left( \frac{t}{\rm Myr} \right)^{1.54} \left( \frac{f_{*}}{0.01} \right) \left( \frac{m_{h}}{10^{8} M_{\odot}} \right).
\end{align}
Again, this expression holds for $t < t_{\rm SN}$. If we wish to expand this calculation to $t > 55$ Myr, we have to recognize that for times $t - t' > 55$ Myr, all the supernovae generated by stars forming at a time $t'$ have exploded already. Thus for $t > 55$ Myr:

\begin{align}
    N(t) = & \ \int_{0}^{t} \dot{N}(t-t') dt' \\
        = & \ \dot{N}(t_{\rm SN}) \times (t - t_{\rm SN}) +  \int_{0}^{t_{\rm SN}} \dot{N}(t-t')dt \\
        = & \ 488  \left( \frac{f_{*}}{0.01} \right) \left( \frac{m_{h}}{10^{8} M_{\odot}} \right) \left(\frac{t - t_{\rm SN}}{{\rm Myr}} \right) + 17237.
\end{align}
And therefore the final, generalized calculation for the number of supernova having occurred after time $t$ is

\begin{equation}
N(t)= 17237 \left( \frac{f_{*}}{0.01} \right) \left( \frac{m_{h}}{10^{8} M_{\odot}} \right)
    \begin{cases}
        \left( t / t_{\rm SN} \right)^{1.54},  & \text{if } t \leq t_{\rm SN} \\
        (t - t_{\rm SN})/t_{\rm char} + 1, & \text{if } t > t_{\rm SN},
    \end{cases} \label{eq:Nt_analytic}
\end{equation}
where $t_{\rm char} = 36$ Myr is a useful characteristic timescale to define.

\begin{figure}
    \centering
    \includegraphics[width=\linewidth]{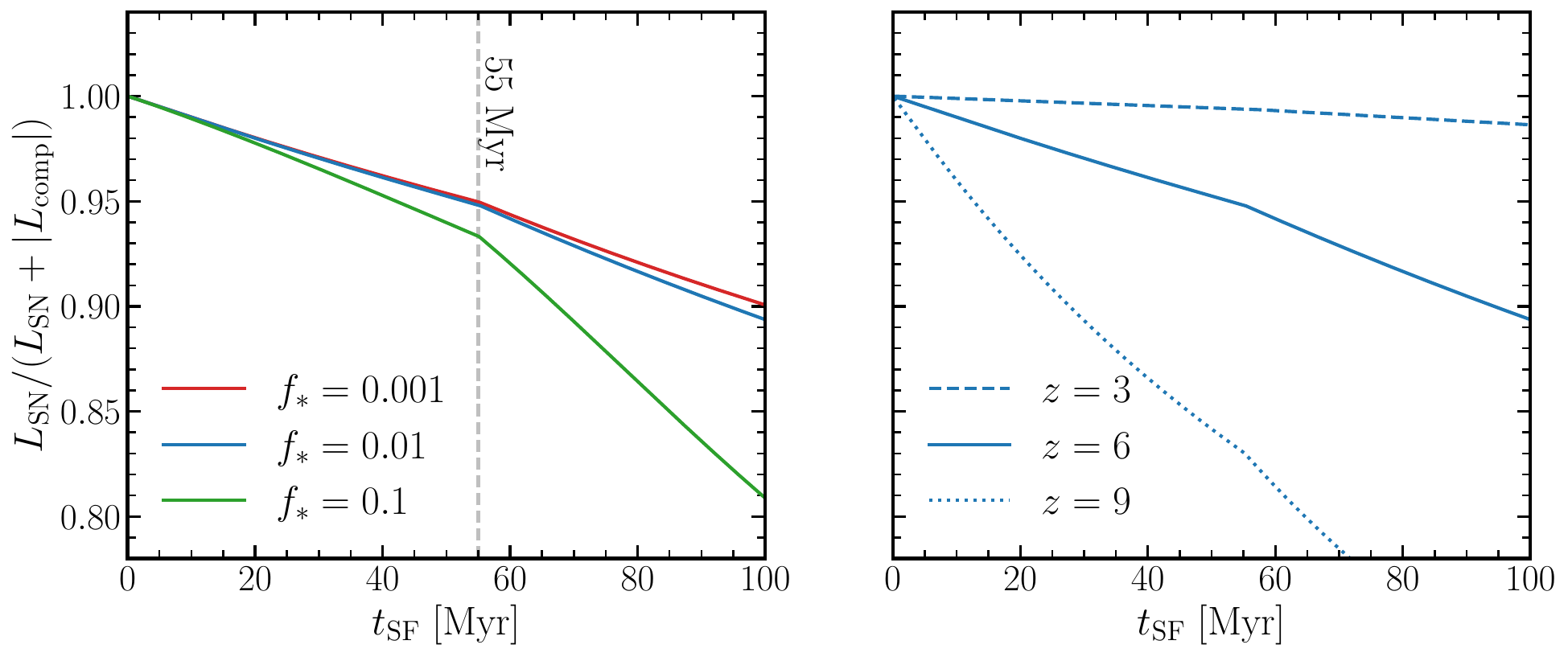}
    \caption{Relative effect of Compton cooling compared to the luminosity from supernova in our model.
    \textit{(Left panel:)} The impact of Compton cooling for multiple $f_{*}$ values at $z = 6$. At the low end of possible $f_{*}$ values, Compton cooling has a maximum effect of $10 \%$ after $t_{\rm SF} = 100$ Myr. The vertical dashed line is $t_{\rm SN} = 55$ Myr (see appendix \ref{sec:N_analytic}).
    \textit{(Right panel:)} Similar to the left panel, but at various redshifts (and with $f_{*} = 0.01$). At lower redshifts, Compton cooling becomes less important, though at early times it cannot be ignored.}
    \label{fig:Lcompton_effect}
\end{figure}

\section{Effect of Compton cooling on the pressure from hot gas} \label{sec:compton_cooling_approx}

In this section, we quantify the effects of Compton cooling by running our galaxy source model (section \ref{sec:source_model}) for $t_{\rm SF} = 100$ Myr and track the luminosity output from supernova and from Compton cooling throughout.

In figure \ref{fig:Lcompton_effect}, we plot the ratio of supernova luminosity to the total luminosity impacting the pressure force for a number of parameters (though we take the absolute value of the luminosity of Compton cooling). In the left panel, we find that our approximation of zero Compton cooling is correct to $\sim 10 \%$ if $f_{*} \lesssim 0.01$
and to $\sim 20 \%$ at higher values of $f_{*}$. We also find that the curves change slope at $t_{\rm SF} \sim 55$ Myr, which aligns with our timescale of $t_{\rm SN} = 55$ Myr (see appendix \ref{sec:N_analytic}). In the right panel of the figure, we plot the effect of Compton cooling for different redshifts. We find that at $z = 3$ and $t_{\rm SF} = 100$ Myr, Compton cooling provides a $\sim 1 \%$ effect on the total pressure luminosity, though at $z = 9$, the Compton cooling quickly becomes important. Therefore, our approximations of zero Compton cooling in section \ref{sec:approx} are accurate for $z \lesssim 6$.

\acknowledgments
We thank Sahil Hegde and Michael Wyatt for useful discussions. This work was supported by NASA through award 80NSSC22K0818 and by the National Science Foundation through award AST-2510939.

This work has made extensive use of NASA's Astrophysics Data System (\href{http://ui.adsabs.harvard.edu/}{http://ui.adsabs.harvard.edu/}) and the arXiv e-Print service (\href{http://arxiv.org}{http://arxiv.org}), as well as the following software: \textsc{matplotlib} \cite{Matplotlib}, \textsc{numpy} \cite{numpy}, \textsc{astropy} \cite{Astropy}, \textsc{scipy} \cite{Scipy}, and \textsc{colossus} \cite{Diemer2018_colossus2}.

\bibliography{bib.bib}{}
\bibliographystyle{JHEP.bst}

\end{document}